\documentclass{aa}

\usepackage{natbib}

\usepackage[section]{algorithm}

\usepackage{algorithmic}
\usepackage{amsmath}
\usepackage{color}
\usepackage{graphicx}
\usepackage{hyperref}
\usepackage{mathcomp}
\usepackage{mathtools}
\usepackage{mdwmath}
\usepackage{mdwtab} 
\usepackage{pifont} 
\usepackage{subfig}
\usepackage{txfonts}
\usepackage{url}                 
\usepackage{graphicx}
\usepackage{amssymb}
\usepackage{epstopdf}

\definecolor{bred}{rgb}{0.9,0,0}
\definecolor{green}{rgb}{0,0.5,0}
\definecolor{linkcolour}{rgb}{0.64,0,0}

\def\simlt{\lower.5ex\hbox{$\; \buildrel < \over \sim \;$}}
\def\simgt{\lower.5ex\hbox{$\; \buildrel > \over \sim \;$}}

\hypersetup{colorlinks,breaklinks,urlcolor=linkcolour, linkcolor=linkcolour}

\bibpunct{(}{)}{;}{a}{}{,}

\date{March 8, 2018} 

\hypersetup{draft}

\begin{document}

\titlerunning{Solving linear equations with messenger-field and conjugate gradient techniques}
\authorrunning{J. Pape\v{z}, L. Grigori, R. Stompor}

\title{Solving linear equations with messenger-field and conjugate gradient techniques -- an application to CMB data analysis. }
\author{J. Pape\v{z}\inst{1}\thanks{\email{Jan.Papez@inria.fr}}, L. Grigori\inst{1}, R. Stompor\inst{2}}
\institute{
INRIA Paris, Sorbonne Universit\'e, Univ Paris-Diderot SPC, CNRS, Laboratoire Jacques-Louis Lions, \'equipe ALPINES, France
\and AstroParticule et Cosmologie, Univ Paris Diderot, CNRS/IN2P3, CEA/Irfu, Obs de Paris, Sorbonne Paris Cit\'{e}, France
}

\abstract{
We discuss linear system solvers invoking a messenger-field and compare them with (preconditioned) conjugate gradient approaches. We show that the messenger-field techniques correspond to fixed point iterations of an appropriately preconditioned initial system of linear equations. We then argue that a conjugate gradient solver applied to the same preconditioned system, or equivalently a preconditioned conjugate gradient solver using the same preconditioner and applied to the original system, will in general ensure
at least a comparable and typically better performance in terms of the number of iterations to convergence and time-to-solution. 
We illustrate our conclusions with two common examples drawn from the cosmic microwave background (CMB) data analysis: Wiener filtering and map-making. In addition, and contrary to the standard lore in the CMB field, we show that the performance of the preconditioned conjugate gradient solver can depend significantly on the starting vector. This observation seems of particular importance in the cases of map-making of high signal-to-noise ratio sky maps and therefore should be of relevance for the next generation of CMB experiments.
}


\maketitle

\section{Introduction.}
\label{sect:intro}

Studies of the cosmic microwave background (CMB) anisotropies have been driving progress in our understanding of the universe for nearly a quarter of a century. The current forefront of the CMB research is the characterization of polarization properties of the CMB anisotropies. The next generation of the CMB observatories has been, and is, designed to ensure that the scientific potential of this new probe is fully exploited. This calls for advanced, high-performance data analysis techniques applicable to enormous data sets which will be collected by these new observatories. 

The analysis of data from CMB observations commonly involves solutions of large, structured linear systems of equations. Two typical and important examples of such systems are map-making and Wiener-filter systems of equations (see, e.g., \cite{Janssen1992, Bunn1994}, respectively, for early pioneering work and \cite{Poletti2017, Seljebotn2017} for examples of more recent applications). These systems are solved either as a stand-alone task or as part of a more involved process, such as a power spectrum estimation, which commonly requires multiple solutions of such systems. In this work we study, from a theoretical and practical perspective, two specific algorithms for solving such systems of equations: a preconditioned conjugate gradient (PCG) approach and a messenger-field (MF) technique. 
Both these approaches have been applied in the context of the applications considered here. Of the two, the PCG approach has been more popular and more broadly used to date. Nevertheless, it has been argued in a number of recent papers~\citep[e.g.,][]{ElsWan13,RamLavWan17,HufNae17} that the messenger-field approach can be highly efficient for both these applications and can deliver performance in some cases exceeding that of some specific PCG approaches, while at the same time being more generally feasible and straightforward to implement and apply~\citep{RamLavWan17, HufNae17,Huff2018}. 
We note that it is the combination of all these features that makes the MF approach potentially attractive.
Indeed, performance of the PCG technique is hinged on a choice of a preconditioner matrix, $\mathbf{M}$, and while very efficient preconditioners can be constructed, in principle outperforming other methods,  typically the construction
quickly becomes difficult and potentially prohibitive from a computational point of view.

Specifically, let us consider a linear system of equations, 
\begin{eqnarray}
\label{eqn:origPblm}
\mathbf{A}\,\mathbf{x} \, = \, \mathbf{b}
,\end{eqnarray}
where the system matrix $\mathbf{A}$ is symmetric and positive definite (SPD). Instead of solving this equation directly, in the PCG approach one solves the preconditioned system,
\begin{eqnarray}
\label{eqn:PrecPblm}
\mathbf{M}^{-1}\,\mathbf{A}\,\mathbf{x} \, = \, \mathbf{M}^{-1}\,\mathbf{b},
\end{eqnarray}
applying the conjugate gradient (CG) technique~\citep{GolVLoBook96}. If the preconditioner is chosen in a way such that $\mathbf{M}^{-1}\,\mathbf{A}$ is better conditioned than the original system matrix $\mathbf{A}$, then the solution can be derived in (often significantly) fewer iterations. Hereafter, we define the condition number $\kappa$ as
\begin{eqnarray}
\kappa(\mathbf{A}) \, \equiv \, {\|\mathbf{A}\|_2} \cdot {\|\mathbf{A}^{-1}\|_2}
\label{eqn:condNumDef}
,\end{eqnarray}
where $\|\mathbf{A}\|_2$ is the spectral norm\footnote{The spectral norm of~$\mathbf{A}$ is equal to the largest singular value of~$\mathbf{A}$ defined as the square root of the largest eigenvalue of $\mathbf{A}^\dagger\mathbf{A}$, where $\dagger$ denotes a Hermitian conjugate. If matrix $\mathbf{A}$ is {\em normal}, i.e., $\mathbf{A}\,\mathbf{A}^\dagger \, = \, \mathbf{A}^\dagger\,\mathbf{A}$, then $\mathbf{A}$ can be diagonalized with the help of a similarity operation employing a unitary matrix and the spectral norm of $\mathbf{A}$ is equal to its largest (in magnitude) eigenvalue and the condition number to the ratio of the largest and the smallest eigenvalues.} of the matrix $\mathbf{A}$.
For a good preconditioner $ \kappa(\mathbf{A}) \gg \kappa(\mathbf{M}^{-1}\,\mathbf{A}) \geq 1$.
A preconditioner is therefore better when its inverse succeeds in capturing more essential features of the inverse system matrix $\mathbf{A}^{-1}$, which cannot be computed directly by assumption. The choice of a preconditioner is a key factor in determining the performance of a PCG solver. There exist both generic and case-specific approaches proposed for their construction. Moreover, for many advanced preconditioners, significant savings in terms of the number of iterations to a solution come at the cost of an overhead related to their construction and/or application on every step of iteration.

While these observations make the method comparison cumbersome and potentially limited to very specific cases and concrete implementations, the question of which class of methods is more promising in ensuring sufficient performance for forthcoming data sets, for example in the context of the CMB field, is of actual practical importance. This is the question we tackle in this work in the context of the MF and PCG solvers.

Our methodology is as follows. We first show that any MF method applied to a linear system involves preconditioning of the original set of equations with a specific preconditioner. Then we argue on theoretical grounds, and later demonstrate using a number of study cases that the corresponding PCG algorithm is at least as efficient as, and often much better than, the MF technique, while featuring similar computational complexity. Combining the MF method with a so-called cooling technique can further improve its performance at least in some cases. Nonetheless, this does not seem to affect the overall assessment at least for the specific cooling prescriptions studied in this paper and motivated by earlier work. Consequently, while the MF technique may still provide an interesting alternative in some specific applications, in general the PCG approach seems more promising and can be a better first choice.
Though we demonstrate our conclusions using specific examples from applications to the Wiener filter and map-making procedures, we expect them to hold more generally.

This paper is organized as follows. In Sect.~\ref{sect:MFsolver} we provide a general discussion of the messenger-field technique as a more general class of solvers and compare it with the PCG solvers as far as its convergence and computational aspects are concerned.
Later, we illustrate the general conclusion of this section with the help of numerical experiments applied to simulated CMB data, involving applications of both these techniques to polarized Wiener filtering (Sect.3) and map-making (Sect.~\ref{sect:mapMake}). We conclude in Sect.~\ref{sect:conclusions}. Some more technical considerations are deferred to the Appendices.

\section{Messenger-field iterative solver.}
\label{sect:MFsolver}

In this section we first present a consistent, general algebraic framework, which encompasses, as specific cases, all implementations of the basic messenger field solver proposed to date in the literature. Subsequently, in Sect.~\ref{sect:coolTech} we describe the cooling technique proposed to improve the performance of the messenger method, and in Sects.~\ref{sect:Conv} and ~\ref{sect:complexity} we discuss the general properties of this broad class of solvers, contrasting them with those of the PCG technique. We develop this general discussion, referring to the specific solvers developed in the context of the applications described in the following sections of this paper.

\subsection{Basic approach}
\label{sect:basicsMF}

Let us consider a system of linear equations as in Eq.~(\ref{eqn:origPblm}).
In general, the messenger-field approach involves a split of the system matrix $\mathbf{A}$, such as
\begin{eqnarray}
\mathbf{A} \equiv \mathbf{C} \, - \, \mathbf{D} = \mathbf{C} \, \big( \mathbf{I} \, - \, \mathbf{C}^{-1}\,\mathbf{D}\big),
\label{eqn:matSplit}
\end{eqnarray}
where  $\mathbf{C}$ is invertible by construction and its inverse is easy to compute. $\mathbf{I}$ is an identity matrix. 
After multiplying Eq.~(\ref{eqn:origPblm}) by $\mathbf{C}^{-1}$ from the left (which corresponds to \emph{preconditioning} the original system), we get the system
\begin{eqnarray}
\big( \mathbf{I} \, - \, \mathbf{C}^{-1}\,\mathbf{D}\big) \, \mathbf{x} \, = \, \mathbf{C}^{-1}\,\mathbf{b}.
\label{eqn:genWithSplit}
\end{eqnarray}
The MF method introduces an extra data object $\mathbf{t}$, the messenger field, which can be defined as
\begin{eqnarray}
\mathbf{t} & \equiv &  \mathbf{D} \, \mathbf{x} \, + \, \mathbf{b},
\label{eqn:mFieldDefGen}
\end{eqnarray}
meaning that Eq.~(\ref{eqn:genWithSplit}) can be represented as
\begin{eqnarray}
\mathbf{t} & = &  \mathbf{D} \, \mathbf{x} \, + \, \mathbf{b},\\
\mathbf{x} & = & \mathbf{C}^{-1}\,\mathbf{t}.
\label{eqn:mFieldGenWithField}
\end{eqnarray}
This can be used to define an iterative scheme
\begin{eqnarray}
\label{eqn:mFieldGenWithFieldIteratedI}
\mathbf{t}^{\left(i+1\right)} & = &  \mathbf{D} \, \mathbf{x}^{\left(i\right)} \, + \, \mathbf{b},\\
\mathbf{x}^{\left(i+1\right)} & = & \mathbf{C}^{-1}\,\mathbf{t}^{\left(i+1\right)}.
\label{eqn:mFieldGenWithFieldIterated}
\end{eqnarray}
We note that the messenger field $\mathbf{t}$ introduced in this way is a dummy object. Therefore, barring some implementational advantages, the equations above are equivalent to a reduced system from which the messenger field has been explicitly eliminated and which can be directly derived from  Eq.~(\ref{eqn:genWithSplit}). This reads,
\begin{eqnarray}
 \mathbf{x} \, = \, \mathbf{C}^{-1}\,\mathbf{D} \, \mathbf{x} \, + \, \mathbf{C}^{-1}\,\mathbf{b},
 \label{eqn:mFieldGenSingle}
\end{eqnarray}
and the corresponding iterative scheme, see also \citep{ElsWan13, HufNae17}, is given by
\begin{eqnarray}
 \mathbf{x}^{\left(i+1\right)} \, = \, \mathbf{C}^{-1}\,\mathbf{D} \, \mathbf{x}^{\left(i\right)} \, + \, \mathbf{C}^{-1}\,\mathbf{b}. 
 \label{eqn:mFieldGenSingleIterated}
\end{eqnarray}
This is a {\em fixed-point iteration} scheme~\citep[e.g.,][]{SaaBook03} and its derivation is analogous to the derivation of the classical iteration methods that also rely on the splitting of the system matrix as in Eq.\eqref{eqn:matSplit}. The Jacobi iterative method takes~$\mathbf{C}$ as the diagonal of~$\mathbf{A}$, while in the Gauss--Seidel method $\mathbf{C}$ is equal to the lower triangular part (including the diagonal) of~$\mathbf{A}$.

We emphasize that whether we choose to implement the single equation version as in Eq.~(\ref{eqn:mFieldGenSingleIterated}) or the double equation one as in Eqs.~(\ref{eqn:mFieldGenWithFieldIteratedI})--(\ref{eqn:mFieldGenWithFieldIterated}), the result will be the same to within numerical precision as in both these cases we solve the same linear system, Eq.~(\ref{eqn:genWithSplit}), performing equivalent iterations. Consequently, the messenger-field approach is a fixed-point iteration technique applied to a preconditioned system in Eq.~(\ref{eqn:genWithSplit}).
However, this equation can be solved using other means, such as for instance a conjugate gradient (CG) approach, which is typically more efficient than the fixed-point iterations (see, e.g., Sect. 5.5 and 2.3 of \citet{LieStrBook13})
Moreover, solving Eq.~(\ref{eqn:genWithSplit}) with the help of the CG technique is equivalent to solving the initial set of equations, Eq.~(\ref{eqn:origPblm}), using a PCG technique with the preconditioner set to $\mathbf{M} \, \equiv \, \mathbf{C}$. In cases when the fixed-point method is expected to converge very efficiently, that is, when $\mathbf{A} \simeq \mathbf{C}$,  the PCG solver will also perform well since
$\mathbf{C}^{-1}\,\mathbf{A} \simeq \mathbf{I}$, a hallmark of a good preconditioner. Similarly, the MF solver  based on the split involving a good preconditioner will likely be efficient. From a computational point of view, both techniques require multiple applications of the inverse preconditioner $\mathbf{M}^{-1}$ to a vector, thus resulting in similar numerical cost.

The main message of this section is that the messenger-field method involves fixed-point iterations  applied to a preconditioned system of linear equations. Its performance is determined by an adopted split of the system matrix, which also defines the preconditioner applied to precondition the initial system. This preconditioner can be used alternately in a PCG solver employed to directly solve  the initial system, and is expected to ensure performance as good as or better than that of the MF technique, as far as the number of iterations as well as time to convergence are concerned.

\subsection{Cooling technique.}
\label{sect:coolTech}

The convergence of the fixed-point method, Eq.~(\ref{eqn:FPerror}), depends on the components of the initial error $\mathbf{x} - \mathbf{x}^{(0)}$ in the invariant subspaces associated with the eigenvalues of $\mathbf{C}^{-1}\,\mathbf{D}$, especially with the dominant (largest) ones. The cooling technique proposed in \citet{ElsWan13} aims at providing, iteratively, a good initial guess $\mathbf{x}^{(0)}$. In the general setting considered above, the cooling technique 
replaces the original problem, Eq.~(\ref{eqn:origPblm}), represented in a split form as in Eq.~(\ref{eqn:matSplit}) by
\begin{equation}
\label{eqn:cooledPblm}
        \mathbf{A}(\lambda)\mathbf{x}(\lambda) = \mathbf{b}(\lambda), \qquad \mathbf{A}(\lambda) = \mathbf{C}(\lambda) - \mathbf{D}(\lambda)
,\end{equation}
where the \emph{cooling parameter} $\lambda$ is defined so that: (a) for $\lambda = 1$, the above problem is equivalent to the original problem in Eqs.~(\ref{eqn:origPblm}) and~(\ref{eqn:matSplit}); and (b) for $\lambda \to \infty$, $\mathbf{D}(\lambda) \to 0$ and $(\mathbf{C}(\lambda))^{-1}\mathbf{D}(\lambda) \to 0$. 
In the cooling method, $\lambda$ is progressively adapted in the course of the iterations with its value gradually decreasing from an initial and typically rather large value down to $1$. While no general prescription is given in the literature, it has been claimed \citep{ElsWan13,HufNae17} that at least in some applications significant gains can be derived as compared to the fixed-point iterations, if the rate of change of $\lambda$ is appropriately tuned.
In general, an MF method combined with the cooling is no longer a {\em fixed-point} method. However, as this is often the case, if $\lambda$ does not change with each iteration but rather is kept constant for some number of iterations before assuming a new value, the iterations for each of the fixed values of the parameters are fixed-point (though not of the original system). In such cases, for large values of $\lambda$ it is expected that an accurate solution of the modified system, Eq.~(\ref{eqn:cooledPblm}), that is, $\mathbf{x}(\lambda)$, can be recovered within a few (fixed-point) iterations. Naturally, $\mathbf{x}(\lambda)$ can be far from the desired solution of the actual system, $\mathbf{x}(1)$, however, it can be a good starting vector for the next round of fixed-point iterations, this time with a smaller value of $\lambda$.  
The relative performance of the cooling method compared to that of the PCG solver of the initial equation, i.e., with $\lambda = 1$, is unclear, and the freedom in defining the rate at which $\lambda$ is changed makes the mathematical analysis of this method difficult; its potential advantages over others are therefore also difficult to anticipate. Consequently, in this work we resort to numerical experiments to investigate the pros and cons of this technique in the specific cases of interest (Sects.~\ref{sect:WF} and~\ref{sect:mapMake}).

We note however that a PCG solver could be used instead of the fixed-point iterations within the cooling scheme.
Though, the fixed-point iterations would still be preferable whenever the value of~$\lambda$ is adjusted after each iteration, or every few iterations, for example, as in the cooling scheme proposed in Sect.~2.2 of \citet{ElsWan13}. However, in the cases when the value of $\lambda$ is kept unchanged over a number of iterations, as in the numerical experiments presented in~\citet{HufNae17} and in \citet{RamLavWan17}, replacing the fixed-point iterations by a PCG method is expected to result in some performance gain accumulated from all the gains obtained from the solutions for a fixed value of $\lambda$.

For clarity, hereafter we use the term `messenger-field method' to denote a method which implements the basic MF algorithm as defined in Sect.~\ref{sect:basicsMF}. Whenever cooling is involved, be it combined with the MF method or the PCG one, we explicitly point this out; for example, we refer to  the `cooled MF' or the `PCG with cooling', and vice versa.

\subsection{Convergence}

\label{sect:Conv}

The convergence properties of the classic, fixed-point iteration methods have been studied extensively in the literature (see, e.g.,
Sections 4.1--4.2 of \citet{SaaBook03}, or  Section~10.1 of \citet{GolVLoBook96}). Given our discussion in Sect.~\ref{sect:basicsMF} those results can be directly applied to the messenger field technique. In particular, from Eqs.~(\ref{eqn:mFieldGenSingle}) and~(\ref{eqn:mFieldGenSingleIterated})~\citep[see also,][]{ElsWan13} the error of the $i$th approximation satisfies the following relation.

 \begin{eqnarray}
 \epsilon^{\left(i\right)} & \equiv &   \mathbf{x} \,-\, \mathbf{x}^{\left(i\right)} \, = \, \mathbf{C}^{-1}\,\mathbf{D} \, \big(\mathbf{x} \,-\, \mathbf{x}^{\left(i-1\right)}\big) 
 \nonumber\\
& \equiv & \mathbf{C}^{-1}\,\mathbf{D} \, \epsilon^{\left(i-1\right)} \,
 =  \big[\mathbf{C}^{-1}\,\mathbf{D}\big]^{i} \, \epsilon^{\left(0\right)}.
 \label{eqn:FPerror}
 \end{eqnarray}
This implies \citep[see, e.g.,][]{GolVLoBook96, SaaBook03} that $\| \epsilon^{\left(i\right)} \|$ converges {asymptotically} to zero as long as the spectral radius of $\mathbf{C}^{-1}\,\mathbf{D}$ is smaller than unity. Here $\|\cdot\|$ denotes the Euclidean norm, and the spectral radius, hereafter denoted by $\rho(\cdot),$ is defined as the largest (in magnitude) eigenvalue of the matrix. This observation generalizes to other norms
given their equivalence on finite dimensional spaces; see Appendix~\ref{sect:MFconverges} for more details.

If matrix $\mathbf{C}^{-1}\,\mathbf{D}$ is also normal, then from~\eqref{eqn:FPerror} it follows that,
\begin{eqnarray}
 \| \epsilon^{\left(i\right)} \| \, \leq \, \rho(\mathbf{C}^{-1}\,\mathbf{D}) \, \| \epsilon^{\left(i-1\right)} \| \,,
 \label{eqn:FPreduction}
\end{eqnarray}
in the Euclidean norm. Therefore, in this case the convergence is not only asymptotic but also monotonic. The normality of $\mathbf{C}^{-1}\,\mathbf{D}$ is also typically necessary~\citep[see, e.g., Sect.~4.1.6 of][for relevant examples]{BjoeBook15}. Consequently, in general some care may need to be exercised in choosing a specific split of the system matrix, Eq.~\eqref{eqn:matSplit}, to ensure that it satisfies both these conditions. This is indeed the case for the Wiener filter application (see,~\citet{ElsWan13} and Appendix~\ref{sect:MFconverges}), as the spectral radius of $\mathbf{C}^{-1}\,\mathbf{D}$ is always smaller than $1$, assuming that the corresponding system matrix~$\mathbf{A}$ is non-singular (see below) and $\mathbf{C}^{-1}\,\mathbf{D}$ is normal. For the map-making application in the rendition of~\citet{HufNae17}, arguments similar to those given in Appendix~\ref{sect:MFconverges} can be used to show that also in this case $\rho(\mathbf{C}^{-1}\,\mathbf{D}) < 1$ for non-singular systems, however the normality of $\mathbf{C}^{-1}\,\mathbf{D}$ remains unclear at this stage.

The character of the convergence will in general depend on the choice of the norm. We show here that it remains monotonic in the $\mathbf{A}$-norm (often called \emph{energy norm)} of the error, if  $\mathbf{C}^{-1}$ and $\mathbf{D}$ are real and symmetric as is indeed the case in the applications studied here. The $\mathbf{A}$-norm is hereafter defined as,
\begin{eqnarray}
\label{eq:energynorm}
        \| \mathbf{x} - \mathbf{x}^{(i)} \|_\mathbf{A} \equiv \left( (\mathbf{x} - \mathbf{x}^{(i)})^T \mathbf{A} (\mathbf{x} - \mathbf{x}^{(i)}) \right)^{1/2}.
\end{eqnarray}
This is one of the norms we use in the follow-up numerical examples.

Using Eq.~\eqref{eqn:FPerror} and $\| \mathbf{v} \|_\mathbf{A} = \| \mathbf{A}^{1/2} \, \mathbf{v} \|$, we obtain,
\begin{equation}
\label{eqn:mondecofAnorm}
        \| \mathbf{x} - \mathbf{x}^{(i)} \|_\mathbf{A} \, \leq \, \| \mathbf{B} \|_2 \cdot \| \mathbf{x} - \mathbf{x}^{(i-1)} \|_\mathbf{A},
\end{equation}
where,
\begin{eqnarray}
\mathbf{B} \, \equiv \, \mathbf{A}^{1/2} \, (\mathbf{C}^{-1} \mathbf{D}) \, \mathbf{A}^{-1/2},
\end{eqnarray}
and $\|\cdot\|_2$ denotes the spectral norm, defined as in Eq.~(\ref{eqn:condNumDef}).

To ensure \emph{monotonic} convergence of the iterative scheme, Eq.~\eqref{eqn:mFieldGenSingleIterated}, in the energy norm it is therefore enough to require that
\begin{eqnarray}
\label{eqn:mondecofAnorm2}
        \| \mathbf{B} \|_2 < 1.
\end{eqnarray}
If $\rho(\mathbf{C}^{-1}\,\mathbf{D}) < 1$, then Eq.~\eqref{eqn:mondecofAnorm2}
is satisfied whenever matrix $\mathbf{B}$ is normal, that is, $\mathbf{B}\,\mathbf{B}^\dagger \, = \, \mathbf{B}^\dagger\,\mathbf{B}$, and therefore it holds that
\begin{eqnarray}
        \| \mathbf{B} \|_2 = \rho\left( \mathbf{B} \right) \, = \, \rho(\,\mathbf{A}^{1/2} \, (\mathbf{C}^{-1} \mathbf{D}) \, \mathbf{A}^{-1/2}) \, = \, \rho\left( \mathbf{C}^{-1} \mathbf{D} \right).
\end{eqnarray}
Here the leftmost equality uses the fact that a normal matrix can be diagonalized using a unitary matrix, and the rightmost follows from the fact that a non-singular similarity transformation, here with $\mathbf{A}^{1/2}$, preserves the eigenvalues.

Assuming that $\mathbf{A}$ is real and symmetric and observing from Eq.~\eqref{eqn:matSplit} that $\mathbf{A} \, (\mathbf{C}^{-1}\, \mathbf{D}) = (\mathbf{D}\,\mathbf{C}^{-1}) \, \mathbf{A}$, we can write,
\begin{eqnarray}
        \mathbf{B}^\dagger\,\mathbf{B}
&=& \mathbf{A}^{-1/2} \, (\mathbf{C}^{-1} \mathbf{D})^\dagger \, \mathbf{A} \,  \mathbf{C}^{-1} \, \mathbf{D} \,\mathbf{A}^{-1/2} \nonumber\\
&=&   \mathbf{A}^{-1/2} \, (\mathbf{C}^{-1} \mathbf{D})^\dagger \, \mathbf{D} \, \mathbf{C}^{-1} \, \mathbf{A}^{1/2},\\
\mathbf{B}\,\mathbf{B}^\dagger & = & 
\mathbf{A}^{1/2} \, (\mathbf{C}^{-1} \mathbf{D}) \, \mathbf{A}^{-1} \,  (\mathbf{C}^{-1} \mathbf{D})^\dagger \, \mathbf{A}^{1/2} \nonumber\\
& = &\mathbf{A}^{-1/2} \, \mathbf{D}\, \mathbf{C}^{-1}  \, (\mathbf{C}^{-1} \mathbf{D})^\dagger \, \mathbf{A}^{1/2},
\end{eqnarray}
and therefore, $\mathbf{B}$ is normal if and only if $(\mathbf{C}^{-1} \mathbf{D})^\dagger \, \mathbf{D} \, \mathbf{C}^{-1} = \mathbf{D}\, \mathbf{C}^{-1}  \, (\mathbf{C}^{-1} \mathbf{D})^\dagger$. This is satisfied, for example, whenever $\mathbf{C}$ and $\mathbf{D}$ are real and symmetric as indeed is the case in the setting of \cite{ElsWan13} and \cite{HufNae17}. 

Consequently, in both these applications we expect monotonic convergence of the MF errors in the energy norm.
This is analogous to the PCG technique where the error is bound to decrease monotonically in the energy norm. This may not however be the case for other norms or the residuals.

The convergence rate of the MF solver is then determined by the eigenspectrum of $\mathbf{C}^{-1}\mathbf{D}$. In particular the eigenmode with the largest eigenvalue, that is, the one closest to 1, will be the slowest to converge.

If the system matrix, $\mathbf{A}$, is singular then
\begin{eqnarray}
\rho(\mathbf{C}^{-1}\,\mathbf{D}) & = & 1.
\end{eqnarray}
This is because
if $\mathbf{x}$ denotes a singular eigenvector of $\mathbf{A}$, that is, $\mathbf{A}\,\mathbf{x} \, = \, 0$, and $\mathbf{x}\,\neq \, 0$ then,
\begin{eqnarray}
\mathbf{A}\,\mathbf{x} & = & \mathbf{C}\,(\mathbf{x} \, - \, \mathbf{C}^{-1}\,\mathbf{D}\,\mathbf{x})\,=\,0,
\end{eqnarray}
and hence
\begin{eqnarray}
\mathbf{C}^{-1}\,\mathbf{D}\,\mathbf{x} & = & \mathbf{x},
\end{eqnarray}
and $\mathbf{x}$ is also an eigenvector of $\mathbf{C}^{-1}\,\mathbf{D}$ but with a unit eigenvalue. In such cases, the convergence of the MF solver will typically stall with the norm of the residuals saturating on a level depending on the right-hand side of the system as well as the initial guess. This behavior is analogous to that of other solvers, such as PCG, and it simply reflects the fact that if $\mathbf{A}$ is singular, then there is no unique solution to the linear system.

We assume from now on that the problem is non-singular and show that the PCG method is typically superior to, and never worse than, the fixed-point method in terms of minimizing the energy norm of the error. We first recall key properties of the (P)CG approach; see, for example, \citet{SaaBook03}, Lemma~6.28.

Let $\mathbf{x}^{(CG, i)}$ be the~$i$th approximation given by the CG method for solving $\mathbf{A} \mathbf{x} = \mathbf{b}$ with the initial guess~$\mathbf{x}^{(0)}$. Subsequently,
\begin{eqnarray}
        \mathbf{x} - \mathbf{x}^{(CG, i)} = \hat{\varphi}_i(\mathbf{A})(\mathbf{x} - \mathbf{x}^{(0)}),
\end{eqnarray}
where $\hat{\varphi}_i$ is a polynomial with $\deg(\hat{\varphi}_i) \leq i$, $\hat{\varphi}_i(0) = 1$, which we write succinctly as $\hat{\varphi}_i \in \mathcal{P}^i_0$, and, 
\begin{eqnarray}
\label{eq:CGminimization}
        \| \hat{\varphi}_i(\mathbf{A})(\mathbf{x} - \mathbf{x}^{(0)}) \|_{\mathbf{A}} = \min_{\hat{\psi}_i \in \mathcal{P}^i_0} {\| \hat{\psi}_i(\mathbf{A})(\mathbf{x} - \mathbf{x}^{(0)}) \|_{\mathbf{A}} }.
\end{eqnarray}
Similarly, when $\mathbf{x}^{(PCG, i)}$ is the~$i$th approximation given by the PCG method for solving the system $\mathbf{A} \mathbf{x} = \mathbf{b}$ preconditioned by $\mathbf{C}$, using the initial guess~$\mathbf{x}^{(0)}$, we have
\begin{eqnarray}
        \mathbf{x} - \mathbf{x}^{(PCG, i)} = \varphi_i(\mathbf{C}^{-1} \mathbf{A})(\mathbf{x} - \mathbf{x}^{(0)}),
\end{eqnarray}
with $\varphi_i \in \mathcal{P}^i_0$ and 
\begin{eqnarray}
\label{eq:PCGminimization}
        \| \varphi_i(\mathbf{C}^{-1} \mathbf{A})(\mathbf{x} - \mathbf{x}^{(0)}) \|_{\mathbf{A}} = \min_{\psi_i \in \mathcal{P}^i_0} {\| \psi_i(\mathbf{C}^{-1} \mathbf{A})(\mathbf{x} - \mathbf{x}^{(0)}) \|_{\mathbf{A}} }.
\end{eqnarray}

Let us now consider the fixed-point method as defined in Eq.~(\ref{eqn:mFieldGenSingleIterated}) assuming the same initial guess, $\mathbf{x}^{(0)}$. From Eqs.~\eqref{eq:PCGminimization} and~\eqref{eqn:FPerror},
\begin{align}
 & \| \mathbf{x} - \mathbf{x}^{(PCG, i)} \|_{\mathbf{A}} = \| \varphi_i(\mathbf{C}^{-1}\mathbf{A})(\mathbf{x} - \mathbf{x}^{(0)}) \|_{\mathbf{A}} \\
    &\qquad \leq \| (\mathbf{C}^{-1} \mathbf{D})^i (\mathbf{x} - \mathbf{x}^{(0)}) \|_{\mathbf{A}} = \| \mathbf{x} - \mathbf{x}^{(i)} \|_{\mathbf{A}},
\end{align}
as $(\mathbf{C}^{-1} \mathbf{D})^i = (\mathbf{I}-\mathbf{C}^{-1}\mathbf{A})^i$ and $\psi_i(x) \equiv (1-x)^i \in \mathcal{P}^i_0$.
This means that, in terms of the energy norm of the error, the PCG method converges at least as fast as the fixed-point method. In practice, one can however expect significantly faster convergence, as suggested by Eq.~\eqref{eq:PCGminimization}.
On the other hand, as emphasized earlier, the performance of the MF solvers can be improved by invoking the cooling technique. 
The convergence of the cooled MF approach is more difficult to study theoretically. Even in the cases when the cooling parameter, $\lambda$, is kept constant over some number of iterations, and the method performs fixed-point iterations within each such interval, these are fixed iterations of the modified, not the original, system and the results concerning the error specified earlier in this section apply only when replacing~$\mathbf{x}$ by the modified solution~$\mathbf{x}(\lambda)$ (and the energy norm~$\| \cdot \|_{\mathbf{A}}$ by ~$\| \cdot \|_{\mathbf{A}(\lambda)}$). The convergence of the iterates to the true solution~$\mathbf{x}$ then should be properly discussed and justified for a particular MF application and/or cooling scheme.
In the absence of theoretical results concerning this last method we assess the relative merits of the different solvers via numerical experiments. This is described in the follow-up sections.

\subsection{Computational complexity.}
\label{sect:complexity}

In actual applications, the computational and memory cost per iteration is often as important as algorithmic efficiency. From this perspective, the fixed-point scheme, Eq.~\eqref{eqn:mFieldGenSingleIterated}, is the cheapest method as it requires an evaluation of $\mathbf{C}^{-1}\mathbf{D}\mathbf{x}^{(i)}$ only once per iteration and storing of only two vectors, $\mathbf{x}^{(i)}$, $\mathbf{C}^{-1}\mathbf{b}$. The PCG method requires more memory needed to store up to four or five vectors, depending on the implementation, and each iteration requires two additional inner products plus some scalar multiplications and vector updates. However, typically, and in particular in the applications considered in this paper, the most time-consuming operations are the multiplications by matrix~$\mathbf{A}$ and by~$\mathbf{C}^{-1}$, rendering these additional costs mostly irrelevant. As an example, in Appendix~\ref{sect:Appendix} we describe an implementation of the PCG algorithm in the context of the Wiener filter that allows  for a single PCG iteration to be performed, with a computational cost comparable to the cost of one fixed-point iteration, Eq.~\eqref{eqn:mFieldGenSingleIterated}.

We can further capitalize on using the PCG method whenever the relative residual or an error measure corresponding to the $\mathbf{A}$-norm of the error need to be frequently evaluated; in the extreme case, at each iteration. The residual $\mathbf{r}^{(i)}$ is updated on each PCG iteration and it is therefore at our disposal; this is not the case for the fixed-point iterations, Eq.~\eqref{eqn:mFieldGenSingleIterated}. Similarly, there is a numerically stable way to evaluate the problem-related error measure corresponding to the $\mathbf{A}$-norm of the error; see also Appendix~\ref{sect:Appendix}. This evaluation involves only scalar quantities that are already at our disposal during the PCG iterations.

We conclude that in terms of time per iteration, both approaches, the MF and the corresponding PCG, are comparable, and therefore the number of iterations to convergence is a sufficient comparison metric.

\section{Application to Wiener filtering}
\label{sect:WF}

\subsection{The problem}
\label{sect:problemWF}

Let us consider a sky map $\mathbf{m}$ composed of a sky signal $\mathbf{s}$ and some noise $\mathbf{n}$ due to our instrument, thus
\begin{eqnarray}
\mathbf{m} \, = \, \mathbf{s} \, + \, \mathbf{n}.
\end{eqnarray}
We assume that the sky signal is Gaussian over an ensemble of sky realizations with zero mean and  known covariance given by $\mathbf{S}$. The noise is also Gaussian with zero mean and the covariance given by $\mathbf{N}$ over the ensemble of noise realizations. We further assume that the noise is uncorrelated and therefore its covariance $\mathbf{N}$ is block-diagonal. The minimum variance estimate of the sky signal, that is, its Wiener filter, is then given by~\citep[e.g.,][]{Bunn1994},
\begin{eqnarray}
\mathbf{s}_{WF} = \big(\mathbf{S}^{-1} + \mathbf{N}^{-1}\big)^{-1}\,\mathbf{N}^{-1}\,\mathbf{m}.
\label{eqn:wienerFilterGen}
\end{eqnarray}
Computing the Wiener filter of the measured map, $\mathbf{m}$, requires an inversion of the system matrix, $\mathbf{S}^{-1} + \mathbf{N}^{-1}$. As modern CMB maps may contain up to many millions of pixels this task can indeed be daunting. This is because in general there is no obvious domain in which both the signal and noise covariances are simultaneously diagonal. Indeed, the signal covariance $\mathbf{S}$ is diagonal in the harmonic domain, where the pixel-domain map $\mathbf{m}$ is described by a vector of coefficients $\mathbf{m}_{\ell m}$ obtained as a result of a spherical harmonic transform applied to the map, while the noise covariance is diagonal in the pixel domain, and only diagonal in the harmonic one if the noise is homogeneous, which is unlikely in practice.
Consequently, a standard way to tackle this problem is to rewrite Eq.~(\ref{eqn:wienerFilterGen}) as a linear set of equations,
\begin{eqnarray}
\big(\mathbf{S}^{-1} + \mathbf{N}^{-1}\big) \, \mathbf{s}_{WF} \, = \,\mathbf{N}^{-1}\,\mathbf{m},
\label{eqn:wienerFilterIter}
\end{eqnarray}
and solve these using some iterative method~\citep[e.g.,][]{SDZ2007}. Both CG and PCG techniques have been applied in this context and while the former was found to show a rather unsatisfactory convergence rate, it was demonstrated that this could be improved significantly albeit with the help of a rather advanced and involved (from the implementation point of view)  preconditioner borrowed from multi-grid techniques~\citep{SDZ2007}.

The MF method originally proposed in this context by~\citet{ElsWan13} involves splitting the noise covariance into homogeneous and inhomogeneous parts by representing $\mathbf{N} \equiv \mathbf{\bar N} + \mathbf{T}$, where $\mathbf{T} = \tau \mathbf{I}$ is a homogeneous part, and $\tau = \min(\mbox{diag}(\mathbf{N}))$. This leads to a split of the system matrix $\mathbf{S}^{-1}\,+\,\mathbf{N}^{-1}$  owing to the fact that,
\begin{eqnarray}
\mathbf{N}^{-1} \, = \, \mathbf{T}^{-1} \,  - \, \mathbf{T}^{-1} \, \big(\mathbf{\bar{N}}^{-1} + \mathbf{T}^{-1}\big)^{-1}\, \mathbf{T}^{-1};
\label{eqn:invSplit}
\end{eqnarray}
\citep[see, e.g.,][p.258]{HigBook02}.
Subsequently, taking
\begin{eqnarray}
\mathbf{C} &\equiv& \mathbf{S}^{-1} \, + \, \mathbf{T}^{-1},\label{eq:WFprec}\\
\mathbf{D} &\equiv& \mathbf{T}^{-1}\,(\mathbf{\bar N}^{-1} \, + \, \mathbf{T}^{-1})^{-1}\,\mathbf{T}^{-1},
\end{eqnarray}
and introducing the messenger field $\mathbf{t}$, Eq.~(\ref{eqn:mFieldDefGen}), we can rewrite Eq.~(\ref{eqn:wienerFilterGen}) in its messenger-field representation, that is,
 \begin{eqnarray}
\begin{array}{r c l}
\medskip
{\displaystyle
 \big(\mathbf{\bar{N}}^{-1} + \mathbf{T}^{-1}\big) \,  \mathbf{t}^{\left(i+1\right)}} & {\displaystyle = } & {\displaystyle \mathbf{T}^{-1} \,  \mathbf{s}_{WF}^{\left(i\right)} \, + \,  \, \mathbf{\bar{N}}^{-1}\,\mathbf{m}}, \\
 {\displaystyle \big(\mathbf{S}^{-1} + \mathbf{T}^{-1}\big) \, \mathbf{s}_{WF}^{\left(i+1\right)}} & {\displaystyle = } & {\displaystyle \mathbf{T}^{-1}\mathbf{t}^{\left(i+1\right)},}
 \end{array}
 \label{eqn:wienerFilterMessFieldIter}
\end{eqnarray}
with the former equation solved in the pixel and the latter in the harmonic domain and with the spherical harmonic transforms used to switch between these domains. These equations are equivalent to Eqs.~(3) and (4) of~\citet{ElsWan13}. Their numerical experiments showed that the solver tended to converge quickly to the solution given the desired precision and therefore the method was proposed as an efficient way to resolve the slow convergence problem of the CG method without the need for potentially complex preconditioners needed for an efficient PCG solver where both these methods should be applied directly to the initial problem, Eq.~(\ref{eqn:wienerFilterIter}).

As argued earlier, Eqs.~(\ref{eqn:wienerFilterMessFieldIter}) are equivalent to a fixed-point iteration solver applied (see Eq.~(\ref{eqn:genWithSplit})),
\begin{eqnarray}  
\Big( \mathbf{I} \, - \, \big(\mathbf{S}^{-1} + \mathbf{T}^{-1}\big)^{-1}\,\mathbf{T}^{-1} \, \big(\mathbf{\bar{N}}^{-1} + \mathbf{T}^{-1}\big)^{-1}\, \mathbf{T}^{-1}\Big) \, \mathbf{s}_{WF} =  \ \ \ \ \ \ \ \ \  \nonumber\\
= \big(\mathbf{S}^{-1} + \mathbf{T}^{-1}\big)^{-1}\,\mathbf{T}^{-1} \, \big(\mathbf{\bar{N}}^{-1} + \mathbf{T}^{-1}\big)^{-1} \,\mathbf{\bar{N}}^{-1}\,\mathbf{m},
\label{eqn:wienerFilterSplit}
\end{eqnarray}
which can be rewritten in an explicitly iterative form as
\begin{eqnarray}
\big(\mathbf{S}^{-1} + \mathbf{T}^{-1}\big) \, \mathbf{s}_{WF}^{\left(i+1\right)} =  \mathbf{T}^{-1} \,  \big(\mathbf{\bar{N}}^{-1} + \mathbf{T}^{-1}\big)^{-1} \, \mathbf{T}^{-1} \,  \mathbf{s}_{WF}^{\left(i\right)} \ \ \ \ \ \ \ \ \ \ \ \ \ \ \nonumber \\
 +  \mathbf{T}^{-1} \, \big(\mathbf{\bar{N}}^{-1} + \mathbf{T}^{-1}\big)^{-1}  \, \mathbf{\bar{N}}^{-1}\,\mathbf{m}.
\label{eqn:wienerFilterSplitIterated}
\end{eqnarray}
In the following section, we compare the performance of different solvers applied to Eqs.~(\ref{eqn:wienerFilterIter}), (\ref{eqn:wienerFilterSplit}), and~(\ref{eqn:wienerFilterSplitIterated}). From the general consideration of the previous section our expectation is that the CG solver applied to Eq.~(\ref{eqn:wienerFilterSplit}), and equivalent to the PCG solution of Eq.~(\ref{eqn:wienerFilterIter}) with $\mathbf{M} \equiv \mathbf{C} = \mathbf{S}^{-1}\,+\,\mathbf{T}^{-1}$, should perform better than the messenger-field solver, Eq.~(\ref{eqn:wienerFilterSplitIterated}).

\begin{figure*}[!ht]
\centering
\includegraphics[width=0.385\linewidth]{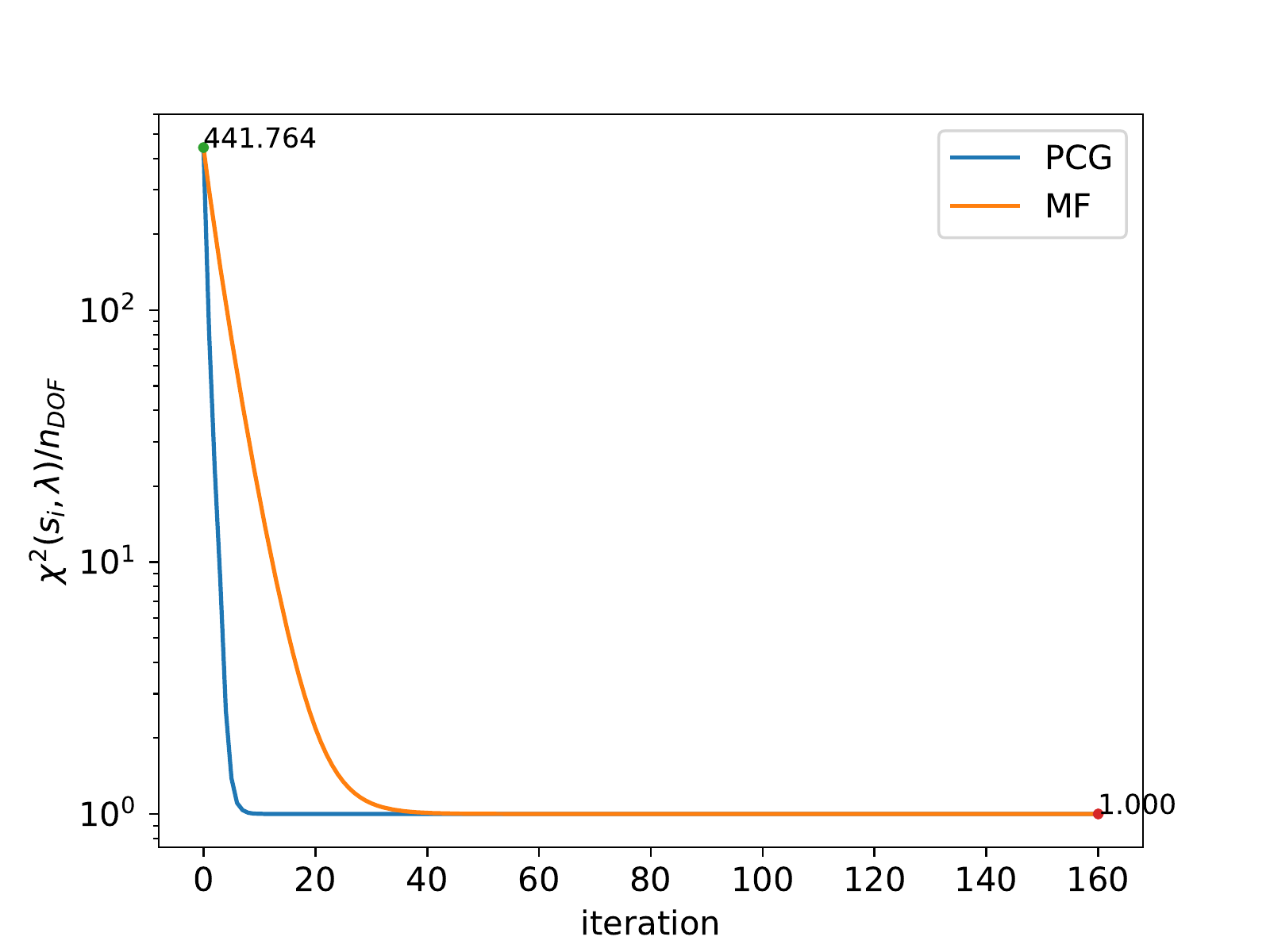}
\includegraphics[width=0.385\linewidth]{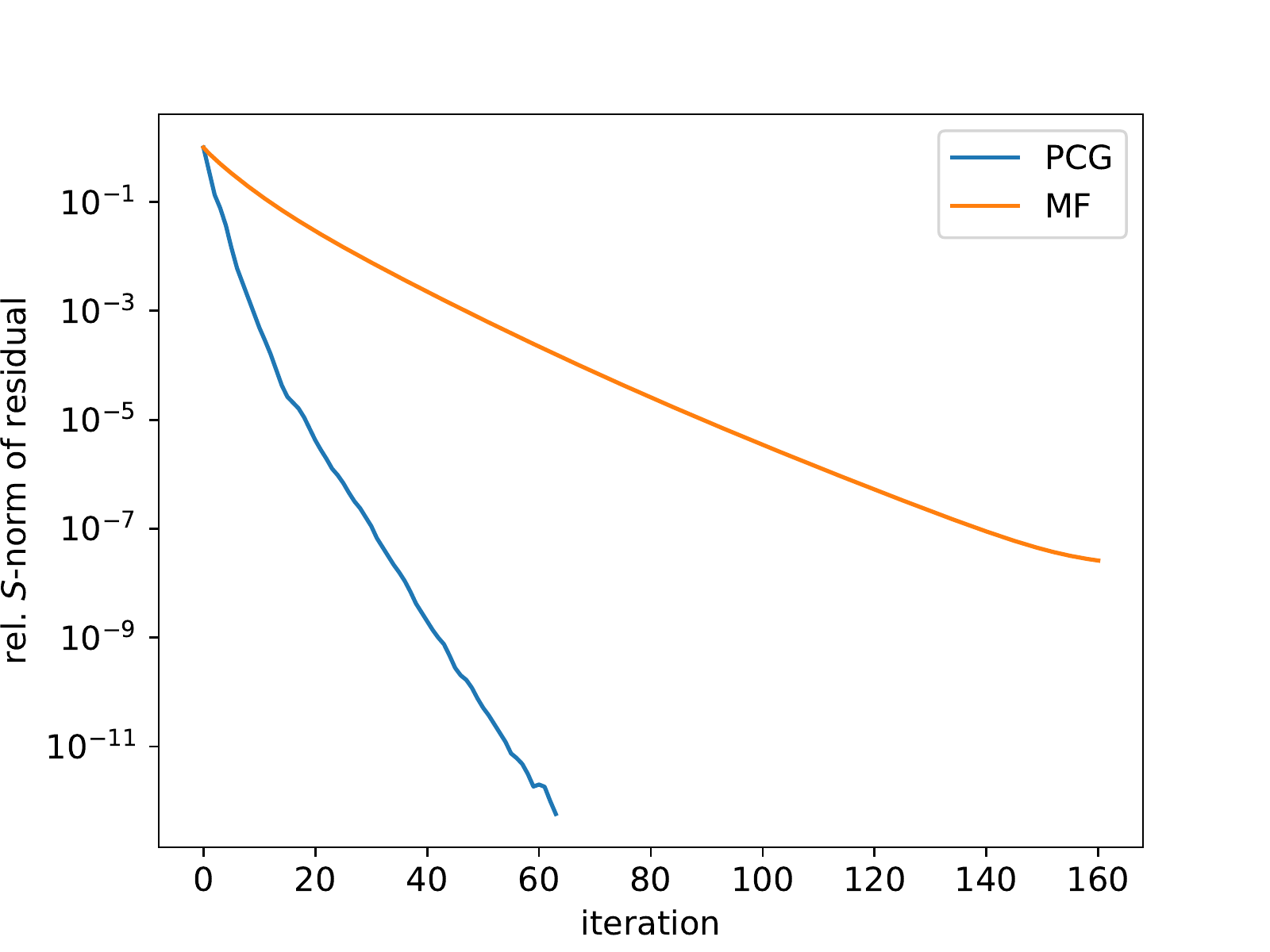}\\
\includegraphics[width=0.385\linewidth]{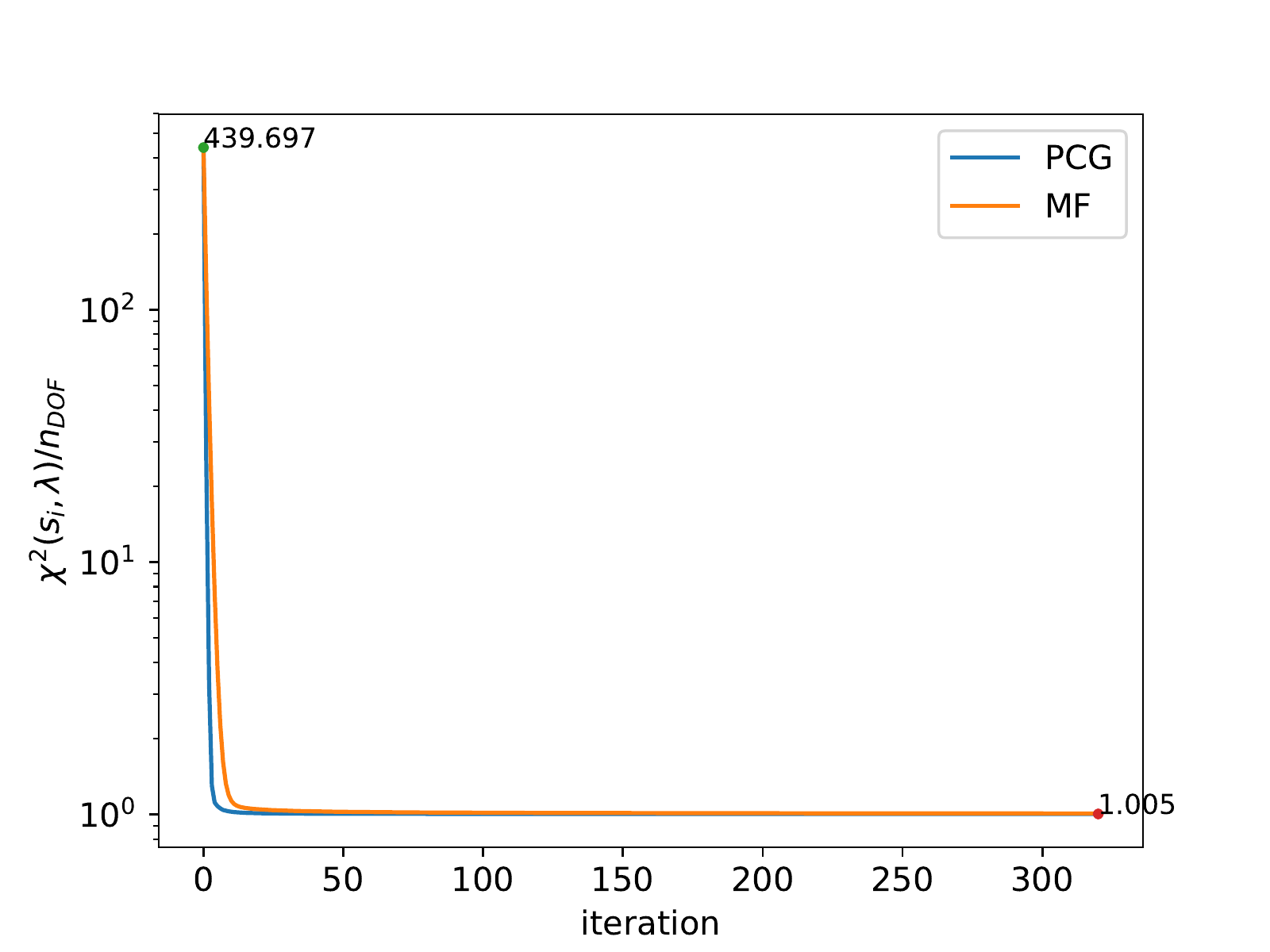}
\includegraphics[width=0.385\linewidth]{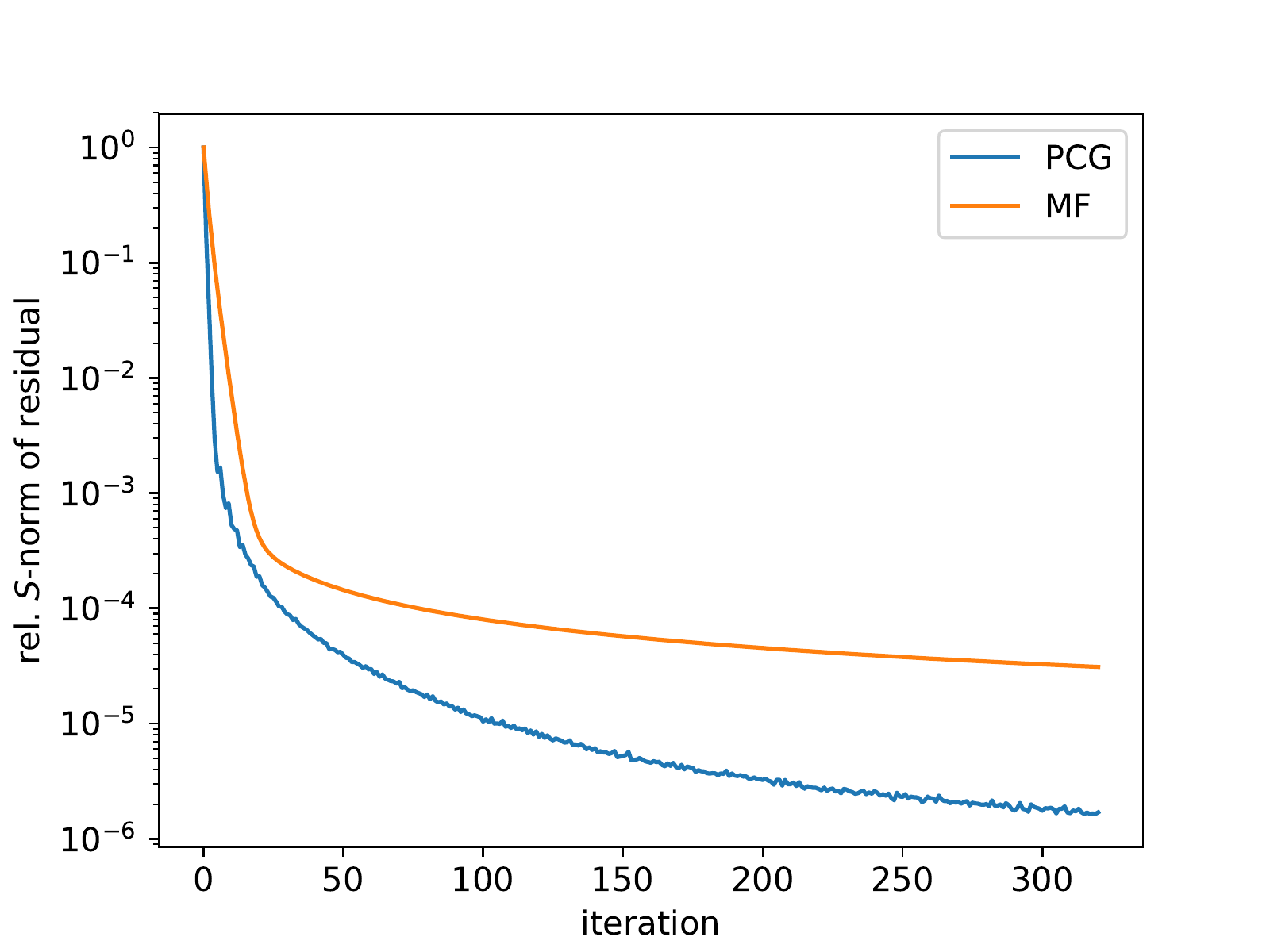}
\caption{Convergence of PCG and MF methods using two different convergence measures: $\chi^2$, Eq.~(\ref{eq:errormeasure}), left panels,  and the $\mathbf{S}$-weighted relative residual, Eq.~(\ref{eq:relSres}), right panels. The top and bottom rows show the case with the full and partial sky coverage, respectively.
\label{fig:FPvsPCG_fsky}
}
\end{figure*}

\subsection{Simulated cases}
\label{sect:WFsims}

To demonstrate and validate our analytical expectation we apply both these solvers to simulated data sets. These are obtained as follows. We first generate maps of three Stokes parameters, $I$, $Q$ and $U$, in the Healpix pixelization~\citep{Gorski2005} with the Healpix resolution parameter $n_{side}$ set to $512$. These maps are computed using a HEALpy routine, {\tt synfast}, providing CMB power spectra  as the input, as computed for the standard cosmological model with the current best values of the parameters~\citep{Planck2015}. In the following calculations we set the band-limit of the sky signal $\ell_{max}$ to $2n_{side}$. This is low enough to ensure the orthogonality of the relevant spherical harmonics over the grid of Healpix pixels. However, it leads to a rank-deficient signal covariance matrix. Consequently, hereafter, its inverse, $\mathbf{S}^{-1}$, is to be understood as a pseudo-inverse. We verified that in selected cases setting $\ell_{max}$ to $3n_{side}$ did not impact our conclusions.
We add to these sky maps inhomogeneous, albeit uncorrelated noise with root mean square (rms) changing over the sky as in the case of the WMAP observations\footnote{We use specifically the noise pattern for the 9-year observation of the V-band, available from {\tt https://lambda.gsfc.nasa.gov/product /map/dr5/maps\_band\_r9\_iqu\_9yr\_get.cfm}.}.

We consider two cases with either full or partial sky observations. In this latter case, only $20$\% of the sky is observed corresponding to the polar cap regions as defined by the Planck HFI mask\footnote{{\tt http://irsa.ipac.caltech.edu/data/Planck/release\_2/ ancillary-data/previews/HFI\_Mask\_GalPlane-apo2\_2048\_R2. 00/index.html}}.

\subsection{Numerical results}
\label{sect:numResWF}

We consider the following solvers.
\begin{itemize}
\item CG applied to the redefined system, Eq.~(\ref{eqn:wienerFilterSplit}), which is equivalent to PCG applied to the original system Eq.~(\ref{eqn:wienerFilterIter}) with a preconditioner given by $\mathbf{M} = \mathbf{S}^{-1}\,+\,\mathbf{T}^{-1}$; (in figures labeled "PCG");
\item MF solver, Eq.~(\ref{eqn:wienerFilterSplitIterated}); (in figures labeled "MF");
\item MF method within three different cooling schemes as proposed in \cite{ElsWan13}, \cite{HufNae17}, and \cite{RamLavWan17}. In the first one, the value of the cooling parameter~$\lambda$ is adjusted adaptively after each iteration. This scheme requires an a~priori knowledge (estimate) on the error measure (see Eq.~(\ref{eq:errormeasure}) below) of the solution. For the purpose of the experiments, this is tightly approximated using the solution of the PCG solver. 
The scheme of \cite{HufNae17} defines a discrete grid of logarithmically spaced values of $\lambda$, which spans the range from $1$ up to some suitable maximal value, which in our runs we set to $\lambda_{max} = 10^4$. For each value of $\lambda$, a fixed number of iterations is performed. 
Though this scheme was suggested specifically for the map-making problem in order to avoid multiple time-consuming reads of the time-ordered data, for the sake of comparison we use it also for the WF experiments. Hereafter, we perform 10~iterations for each of the $16$ values of $\lambda$ and refer to this scheme as "$16\times 10$". For the case with partial sky observations, we continue with the fixed-point iterations~(\ref{eqn:wienerFilterSplitIterated}) for $\lambda=1$.
The cooling scheme of~\cite[Algorithm~1]{RamLavWan17} reduces $\lambda$ by a constant factor, $\eta$, so $\lambda \rightarrow \lambda\times\eta$ and iterates as long as two consecutive approximations satisfy $\| \mathbf{s}^{(i)} - \mathbf{s}^{(i-1)}\| / \| \mathbf{s}^{(i)}\| < \epsilon$. In our experiments we start with $\lambda_{max} = 10^4$, and we set $\eta = 3/4$ and $\epsilon \equiv 10^{-4}$.
\end{itemize}
We start the iterations with a vector of zeros as an initial guess.

\begin{figure*}[!ht]
\includegraphics[width=0.365\linewidth]{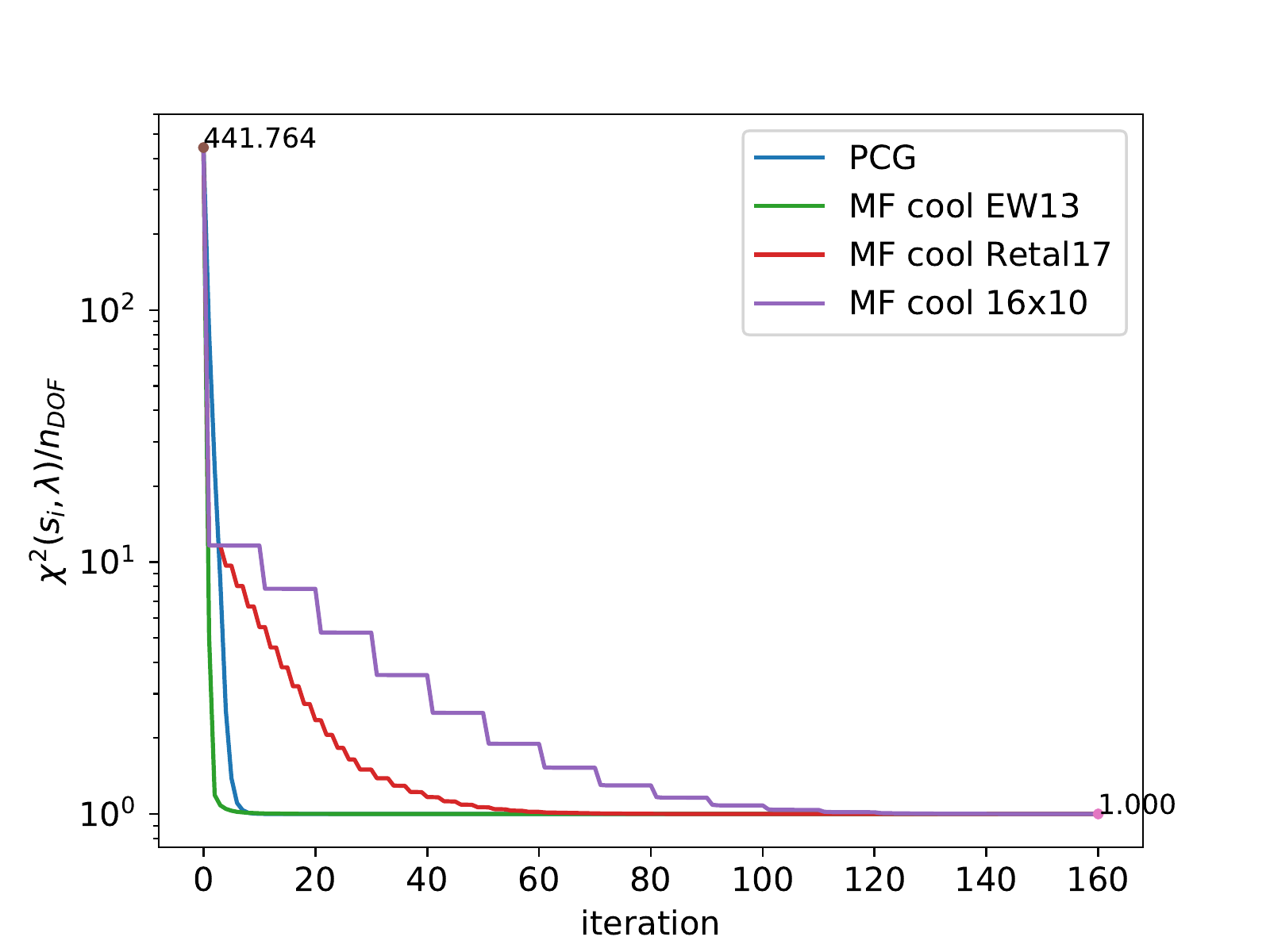}
\hskip -0.6truecm
\includegraphics[width=0.365\linewidth]{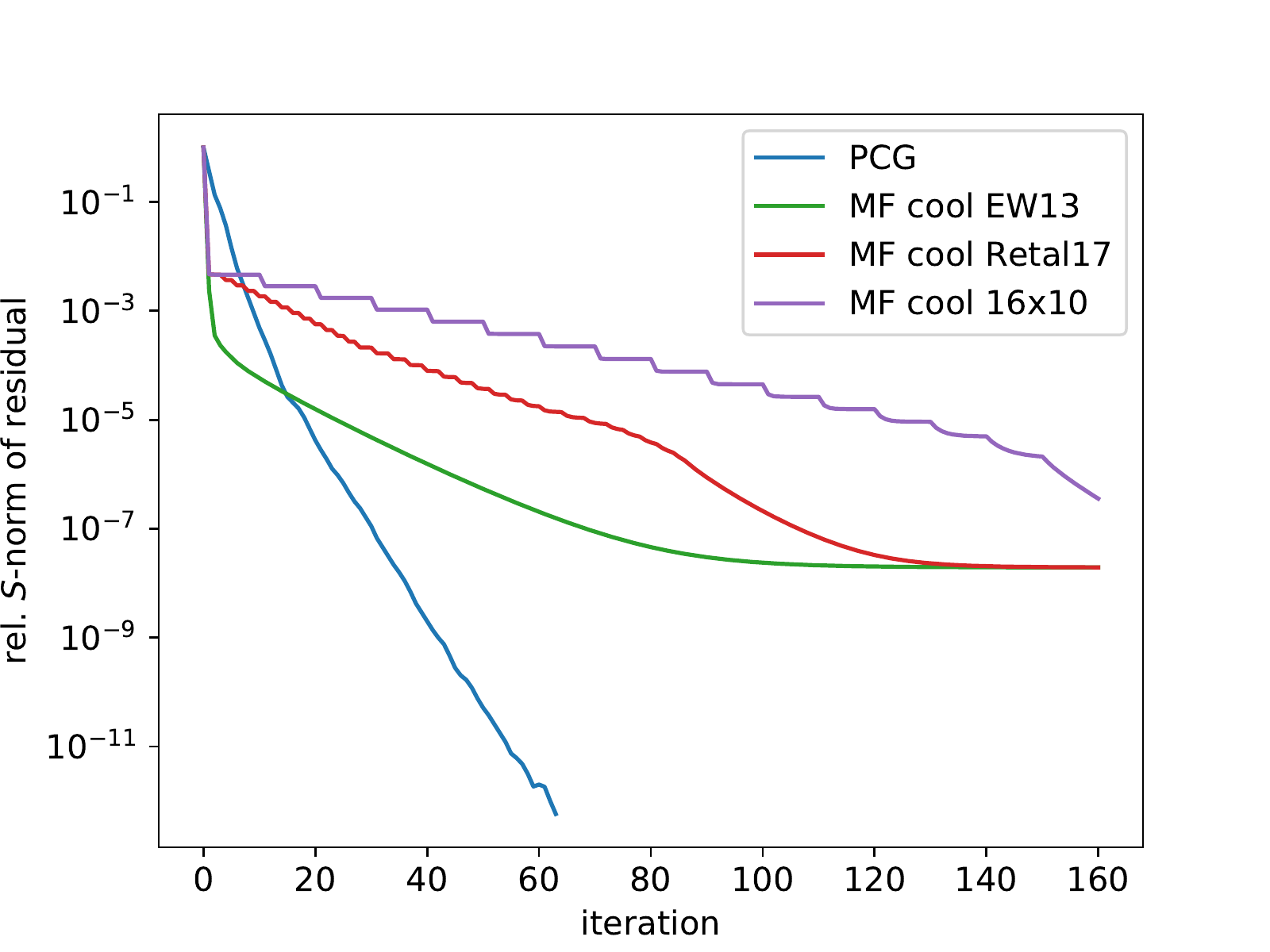}
\hskip -0.6truecm
\includegraphics[width=0.365\linewidth]{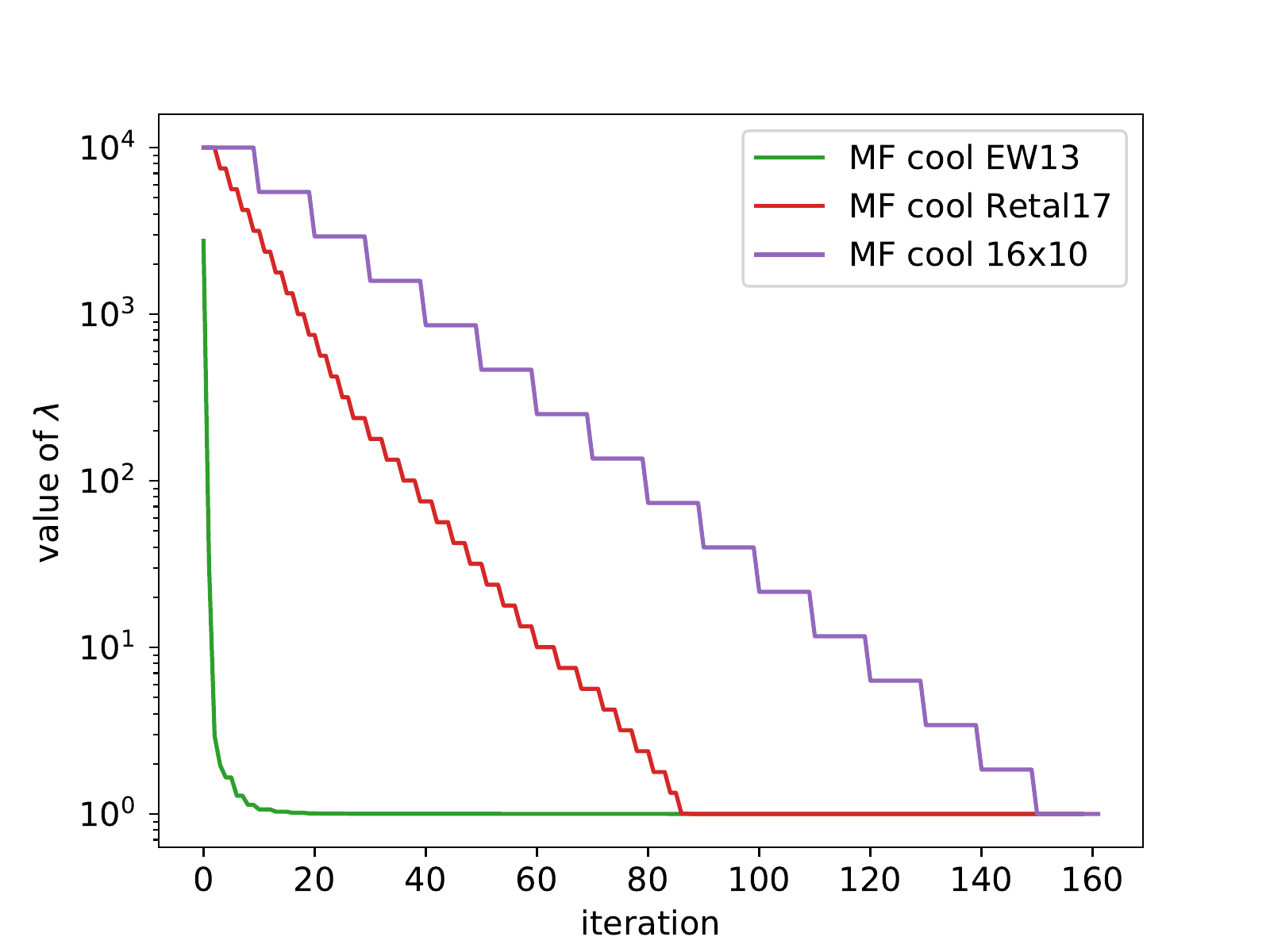}
\caption{Comparison of the convergence of the PCG solver and the MF technique with an adaptive cooling for different cooling prescriptions and using different convergence criteria: the $\chi^2$, left, the $\mathbf{S}$-weighted relative residual, middle. The right panel shows the values of $\lambda$ as a function of the iteration as adapted by the different cooling schemes. These results are for the data sets with the full sky coverage.
}
\label{fig:PCGvsFPcool_fsky1}
\end{figure*}

\begin{figure*}[!ht]
\includegraphics[width=0.365\linewidth]{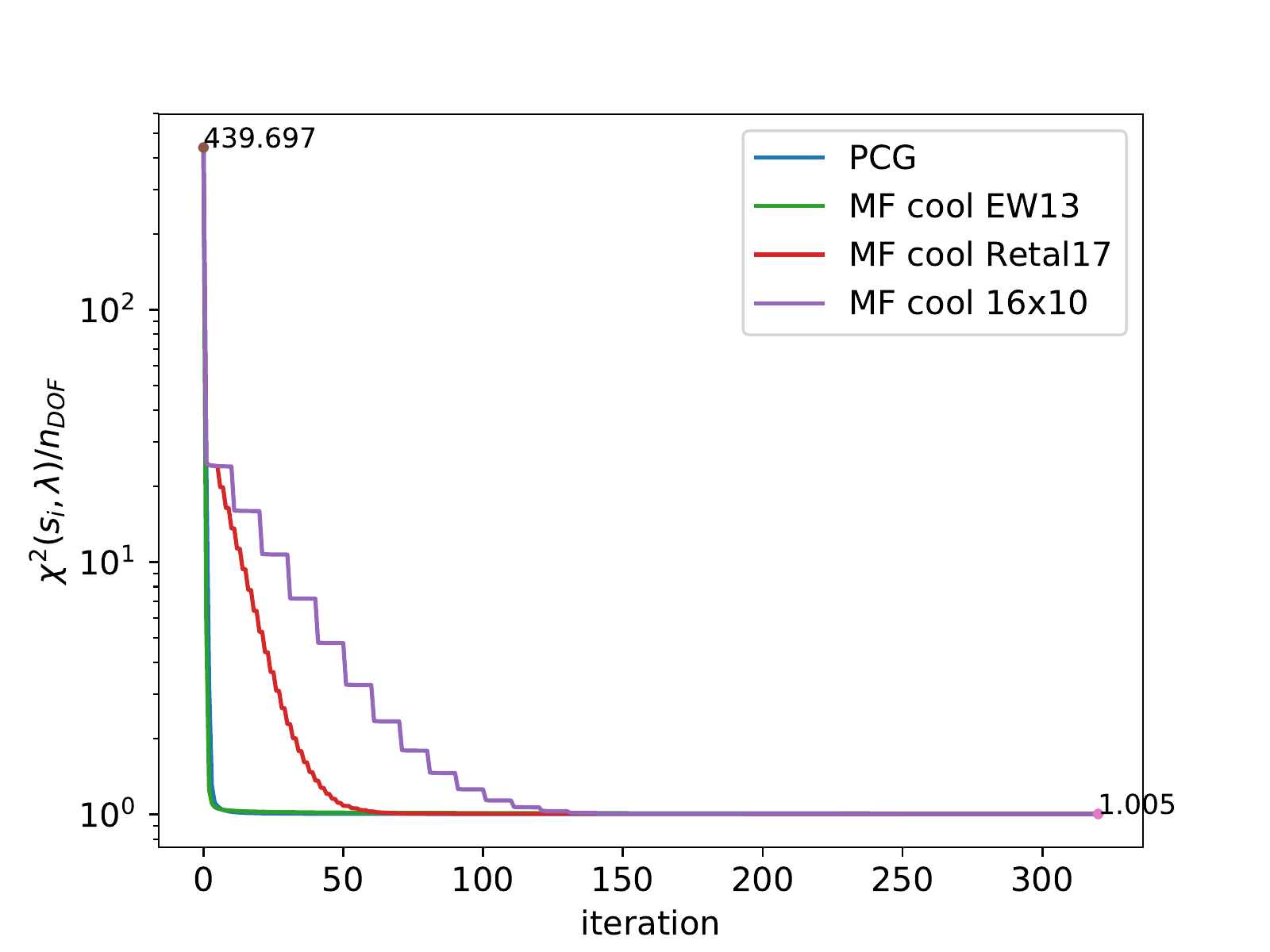}
\hskip -0.6truecm
\includegraphics[width=0.365\linewidth]{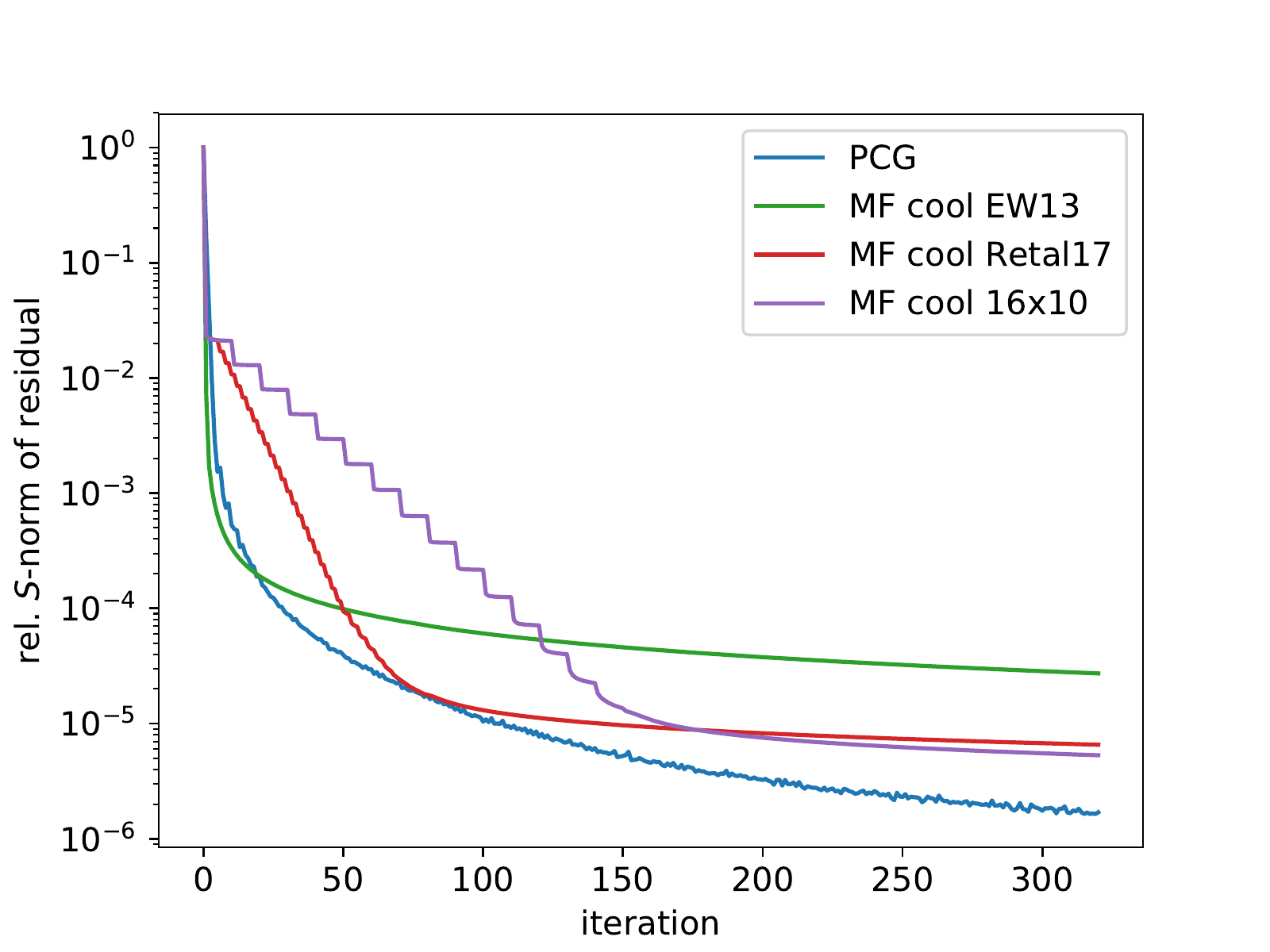}
\hskip -0.6truecm
\includegraphics[width=0.365\linewidth]{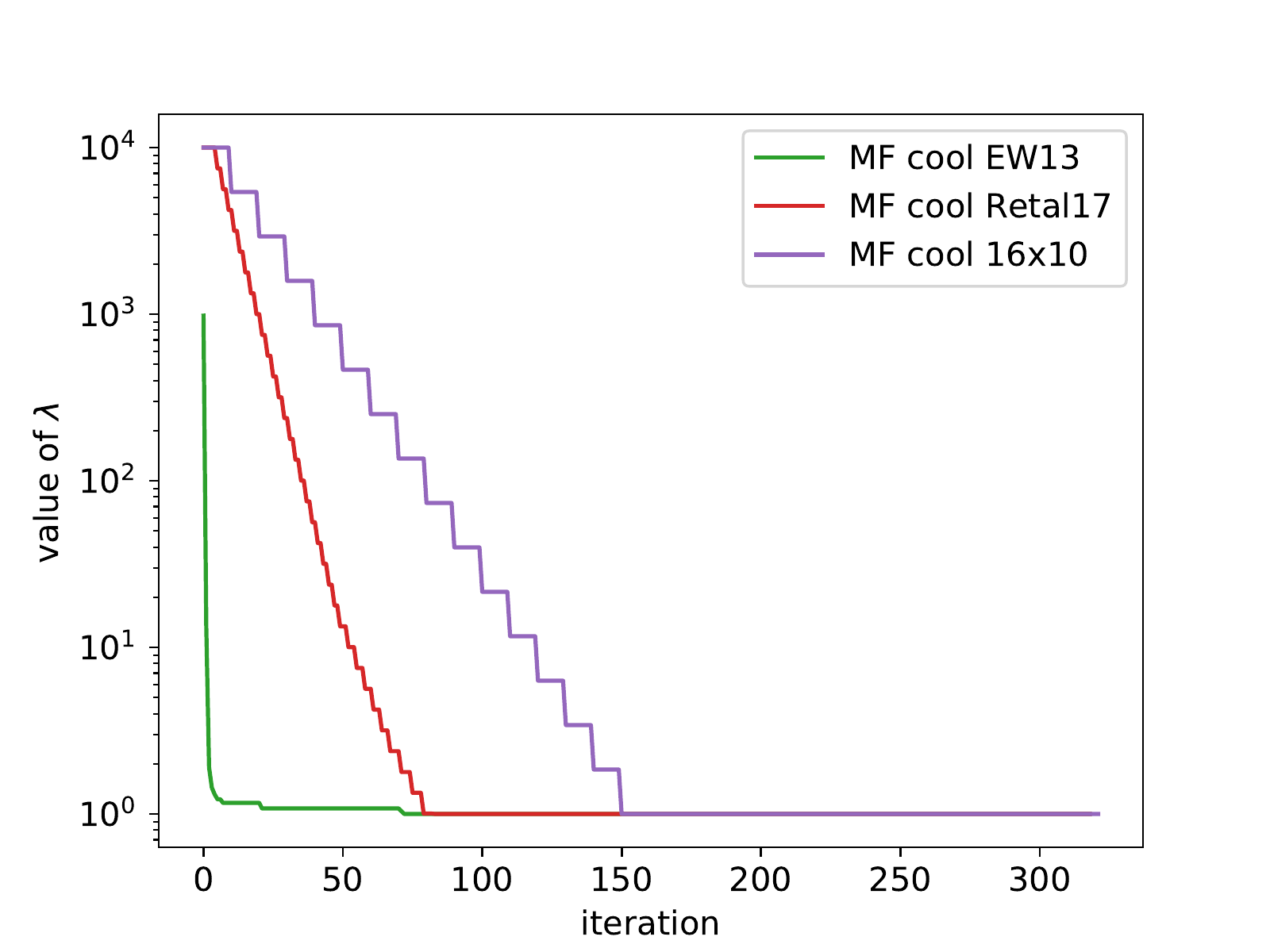}
\caption{As in Fig.~\ref{fig:PCGvsFPcool_fsky1} but for the data set with the partial sky coverage, $f_{sky} = 0.2$.
}
\clearpage
\label{fig:PCGvsFPcool_fsky02}
\end{figure*}

The signal covariance, $\mathbf{S}$, is computed assuming the CMB power spectra as used for the simulations. The noise covariance is block-diagonal in the pixel domain but not proportional to $\mathbf{I}$ as the noise is assumed to be inhomogeneous. It is taken to be exactly the same as the noise covariance used for the simulations.

For all solvers, the inverse signal covariance is applied to a map-length pixel domain vector in the harmonic domain, when first the vector is represented by a vector of its harmonic coefficients computed with the help of a HEALpy routine, {\tt map2alm}; these are subsequently weighted by the inverse of the power spectra and transformed back to the pixel domain using HEALpy's {\tt alm2map} routine. The inverse noise covariance is applied to any pixel-domain vector directly with the elements corresponding to unobserved pixels set to zero.

We note that we always estimate the Wiener-filtered sky signal over the full sky. In all cases shown below we apply all the solvers to exactly the same input data sets.

We have validated our implementations by considering a simplified data set with white noise. 
In this case, the noise covariance $\mathbf{N}$ is proportional to the unit matrix, and the PCG solver with preconditioner $\mathbf{M} = \mathbf{S}^{-1}\,+\,\mathbf{T}^{-1}$, and the MF solver converged to within the numerical precision in a single step as expected, as in this case $\mathbf{T} = \mathbf{N}$.
Moreover, in the cases of the actual simulated data sets used in our test, the results obtained with the different solvers are consistent.

\subsubsection{Convergence metric}
\label{sect:WFmetric}
 
The Wiener-filter problem, Eq.~\eqref{eqn:wienerFilterIter}, can be recast as a minimization of the functional
\begin{equation}
\label{eq:errormeasure}
        \chi^2(\mathbf{x}) = \mathbf{x}^t\mathbf{S}^{-1}\mathbf{x} + (\mathbf{m-x})^t\mathbf{N}^{-1}(\mathbf{m-x}).
\end{equation}
Indeed, we have
\[
        \mbox{argmin } \chi^2(\mathbf{x}) 
        = {\big( \mathbf{S}^{-1} + \mathbf{N}^{-1} \big)}^{-1} \mathbf{N}^{-1} \mathbf{m} = \mathbf{s}_{WF}.
\]
Algebraic manipulations show that $\chi^2$ is directly related to the energy norm, as we have
\[
        \| \mathbf{s}_{WF} - \mathbf{x} \|^2_{\mathbf{S}^{-1} + \mathbf{N}^{-1}} = \chi^2(\mathbf{x}) - \chi^2(\mathbf{s}_{WF}).
\]
We can therefore use physically motivated $\chi^2$ as a convergence measure instead of the energy norm. We note that we expect that 
\begin{eqnarray}
\chi^2(\mathbf{x}) \rightarrow \chi^2(\mathbf{s}_{WF}) = \mathbf{m}^t\,(\mathbf{S}+\mathbf{N})^{-1}\,\mathbf{m}.
\end{eqnarray}
As with the energy norm, Eq.~(\ref{eq:energynorm}), this asymptotic value of $\chi^2$ cannot be straightforwardly computed without knowing the Wiener filter estimate precisely. However, we expect that it should be close to $\langle \chi^2(\mathbf{s}_{WF}) \rangle = n_{Stokes}\,n_{pix} \equiv n_{DOF}$ within a small scatter on order of ${\mathcal O}(\sqrt{2\,n_{Stokes}\,n_{pix}})$, if our assumptions about the sky signal and the noise are correct. Here the angle brackets denote the average over an ensemble of sky and noise realizations and $n_{Stokes}$ stands for the number of considered Stokes parameters and is equal to 3 for most of our tests. We can therefore define the convergence in this case by requiring that the incremental change of $\chi^2$ between consecutive iterations is not larger than some small fraction of $\langle \chi^2(\mathbf{s}_{WF}) \rangle$~\citep[][]{ElsWan13}. If the absolute value of the final $\chi^2$ is statistically inconsistent with the expected one, this could be an indication of prematurely stalled convergence or of a problem with the model assumed for the measured data, $\mathbf{m}$.

Given the discussion of Sect.~\ref{sect:Conv}, we expect that in terms of minimizing the $\chi^2$-measure, the PCG method with preconditioner $\mathbf{M} = \mathbf{S}^{-1} + \mathbf{T}^{-1}$ should be superior to the fixed-point iterations, Eq.~\eqref{eqn:wienerFilterSplitIterated}.

In addition to the $\chi$-measure we also plot the norm of the residual corresponding to the (preconditioned) problem as suggested in \citet{RamLavWan17}. This is given by,
\begin{eqnarray}
        \mathbf{S}^{1/2} \big( \mathbf{S}^{-1} + \mathbf{N}^{-1} \big) \, \mathbf{S}^{1/2}  \mathbf{S}^{-1/2} \mathbf{s}_{WF} = \mathbf{S}^{1/2}  \mathbf{N}^{-1} \mathbf{m};
\end{eqnarray}
~\citep[][Appendix~C]{RamLavWan17}. This system is significantly better conditioned than the original one, Eq.~\eqref{eqn:wienerFilterIter}. The corresponding relative norm of the residual then reads
\begin{equation}
\label{eq:relSres}
        \frac{\|  \mathbf{S}^{1/2} \mathbf{N}^{-1} \mathbf{m} - \mathbf{S}^{1/2} \big( \mathbf{S}^{-1} + \mathbf{N}^{-1}\big) \mathbf{x} \|}
    {\| \mathbf{S}^{1/2} \mathbf{N}^{-1} \mathbf{m}\|} = \frac{\| \mathbf{N}^{-1} \mathbf{m} - \big( \mathbf{S}^{-1} + \mathbf{N}^{-1}\big) \mathbf{x} \|_{\mathbf{S}}}
    {\|\mathbf{N}^{-1} \mathbf{m}\|_{\mathbf{S}}}.
\end{equation} 

\subsubsection{Performance}

Figure~\ref{fig:FPvsPCG_fsky} shows a comparison between the PCG (Eq.~(\ref{eqn:wienerFilterSplit})) and the MF 
(Eq.~(\ref{eqn:wienerFilterSplitIterated})) solvers as applied to the Wiener-filter problem. As expected, PCG indeed reduces the error significantly faster. 

In Figs.~\ref{fig:PCGvsFPcool_fsky1} and~\ref{fig:PCGvsFPcool_fsky02}, we compare PCG (with the MF preconditioner $\mathbf{S}^{-1}\,+\,\mathbf{T}^{-1}$ and $\lambda=1$) with the MF method using the adaptive cooling schemes described above. 
We can see that PCG yields robust performance in all the test cases. 
In the case with full sky observations, the MF solvers (with or without cooling) reach their asymptotic convergence rate and exhibit a plateau of convergence on the level $10^{-7}$ of the relative $\mathbf{S}$-norm of the residual. This is not the case for the PCG solver, which converges to the machine precision level. In the experiment with partial sky coverage, we observe a decrease of the convergence rate for the PCG as well as the MF solvers due to significantly worse conditioning of the problem. However, even in this case, the PCG method is superior to the MF solvers. We expect that using more advanced preconditioners, which can alleviate the effect of very small eigenvalues, can bring a further significant improvement. 

We note, however, that PCG appears to be outperformed by the MF method with the \citet{ElsWan13} cooling proposal  within the first ten or so iterations. As both methods solve different linear systems in this latter case, due to different values of $\lambda$, this does not contradict our conclusions in Sect.~\ref{sect:Conv}. This also does not change the overall assessment of the relative merits of both these techniques as no convergence is then ever reached in terms of any of the considered metrics. We discuss a possible origin of this behavior in Sect.~\ref{sect:numResMapMake}, in the map-making context, where we suggest a simple antidote that could potentially further improve the performance of the PCG approach.

\section{Application to map making}
\label{sect:mapMake}

\subsection{The problem}
\label{sect:problemMapMake}

Data collected by modern, single-dish CMB experiments are modeled as
\begin{eqnarray}
\mathbf{d} = \, \mathbf{P}\,\mathbf{m} \, + \, \mathbf{n},
\label{eqn:dataModel}
\end{eqnarray}
where $\mathbf{d}$ stands for a vector of all measurements, $\mathbf{m}$ is a pixelized map of the sky signal and $\mathbf{n}$ is the instrumental noise. We assume for simplicity that experimental beams are axially symmetric and that the sky signal $\mathbf{m}$ is already convolved with the beam. In this case, pointing matrix $\mathbf{P}$ simply defines which pixel of the map, $\mathbf{m}$, is observed at each measurement and with what weight it contributes to the measurement. In such cases, the pointing matrix is very sparse as it contains only one non-zero element per row for the total intensity measurements, or three for the polarization-sensitive ones. Moreover, $\mathbf{P}^{\, t}\,\mathbf{P}$ is either diagonal or block-diagonal with $3\times3$ blocks. If we assume that the instrumental noise is Gaussian with the covariance given by $\mathbf{N}$, a maximum-likelihood estimate of the sky signal can be written as
\begin{eqnarray}
\mathbf{m}_{ML} & \equiv & \big(\mathbf{P}^{\, t} \, \mathbf{N}^{-1}\, \mathbf{P}\big)^{-1}\, \mathbf{P}^{\, t} \, \mathbf{N}^{-1}\,\mathbf{d},
\label{eqn:mapMake}
\end{eqnarray}
and therefore requires a solution of large linear system. The sizes of the involved object vary significantly, depending on the experiment, but the number of pixels in the map $\mathbf{m}$ can easily reach ${\cal O}(10^6)$, while the number of measurements, ${\cal O}(10^{12-15})$. Consequently, the system can only be solved iteratively, explicitly capitalizing on the structure and sparsity of the involved data objects.

Traditionally~\citep[e.g.,][]{deGasperis2005, Cantalupo2010} the iterative method of choice was a PCG technique with a simple preconditioner given by
\begin{eqnarray}
\label{eq:MMstandardprec}
\mathbf{M} \, = \, \mathbf{P}^{\, t} \, \mbox{diag} (\mathbf{N}^{-1})\, \mathbf{P}.
\end{eqnarray}
Hereafter we refer to this standard preconditioner as block-diagonal or Jacobi. However,
more involved preconditioners have also been considered and found to be successful~\citep[e.g.,][]{Grigori2012, Szydlarski2014, Naes2014, Puglisi2018}. 

\begin{figure*}[!ht]
\centering
\hskip -0.1truecm
\includegraphics[width=0.335\linewidth]{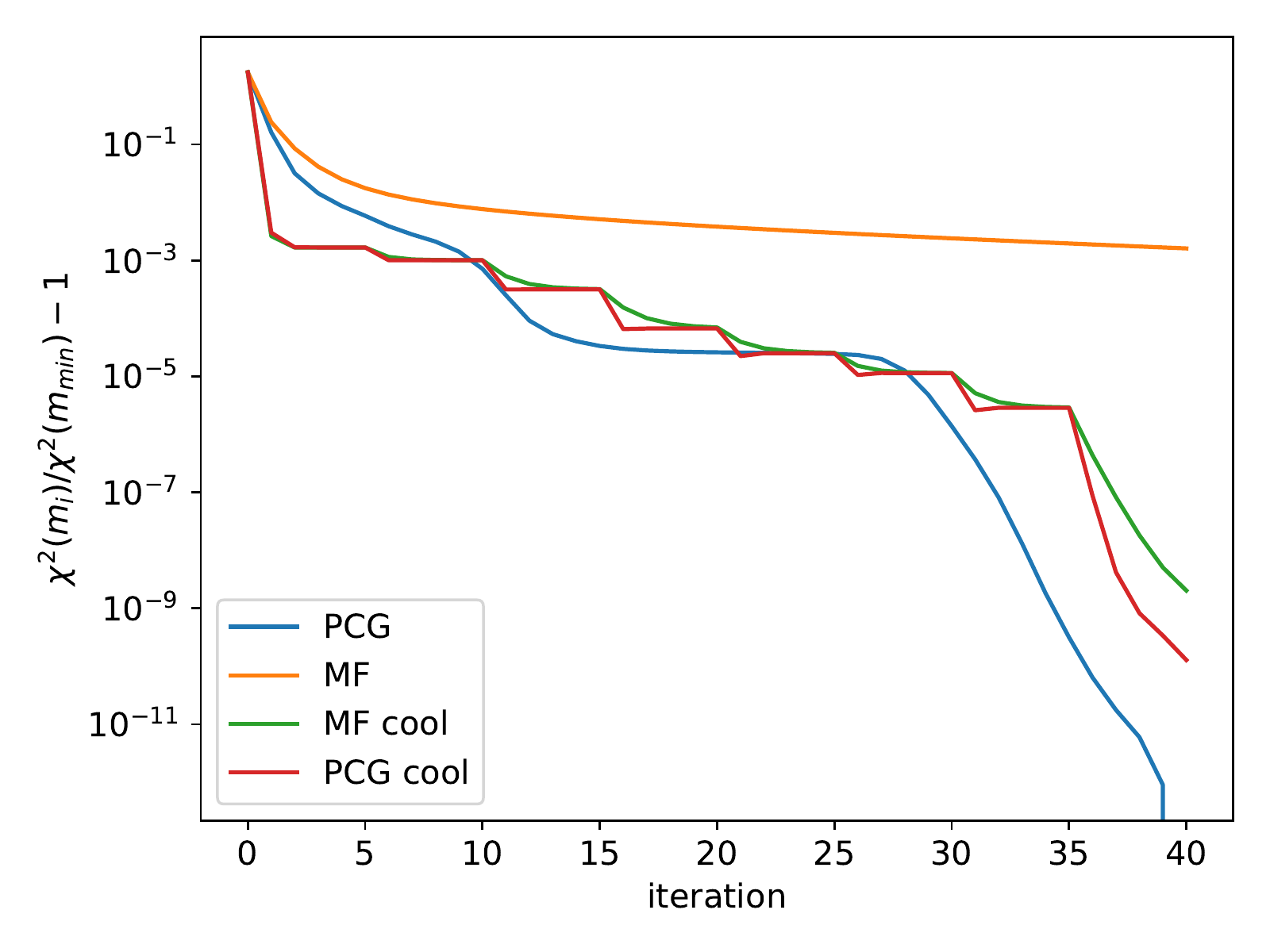}
\hskip -0.1truecm
\includegraphics[width=0.335\linewidth]{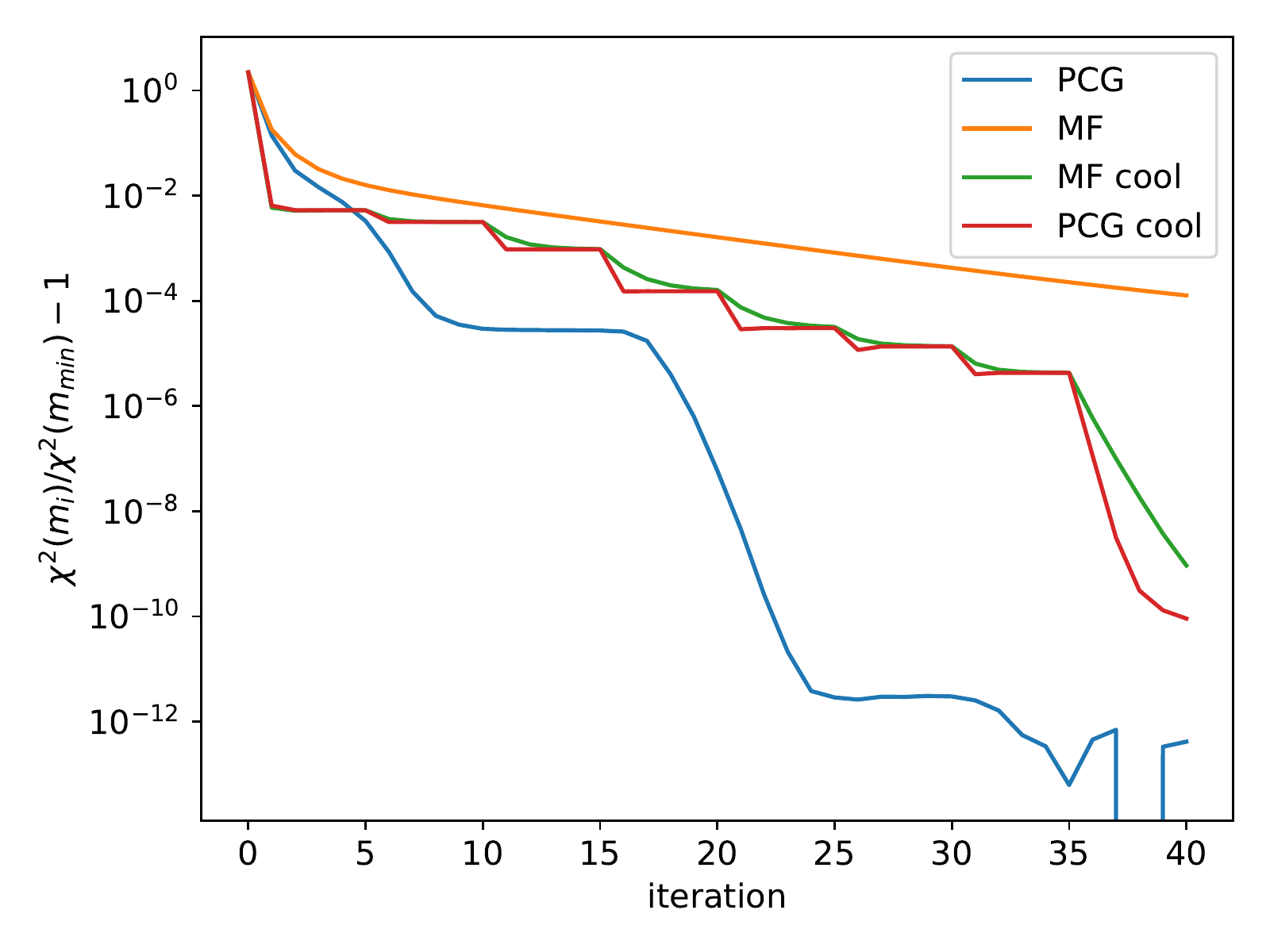}
\hskip -0.1truecm
\includegraphics[width=0.335\linewidth]{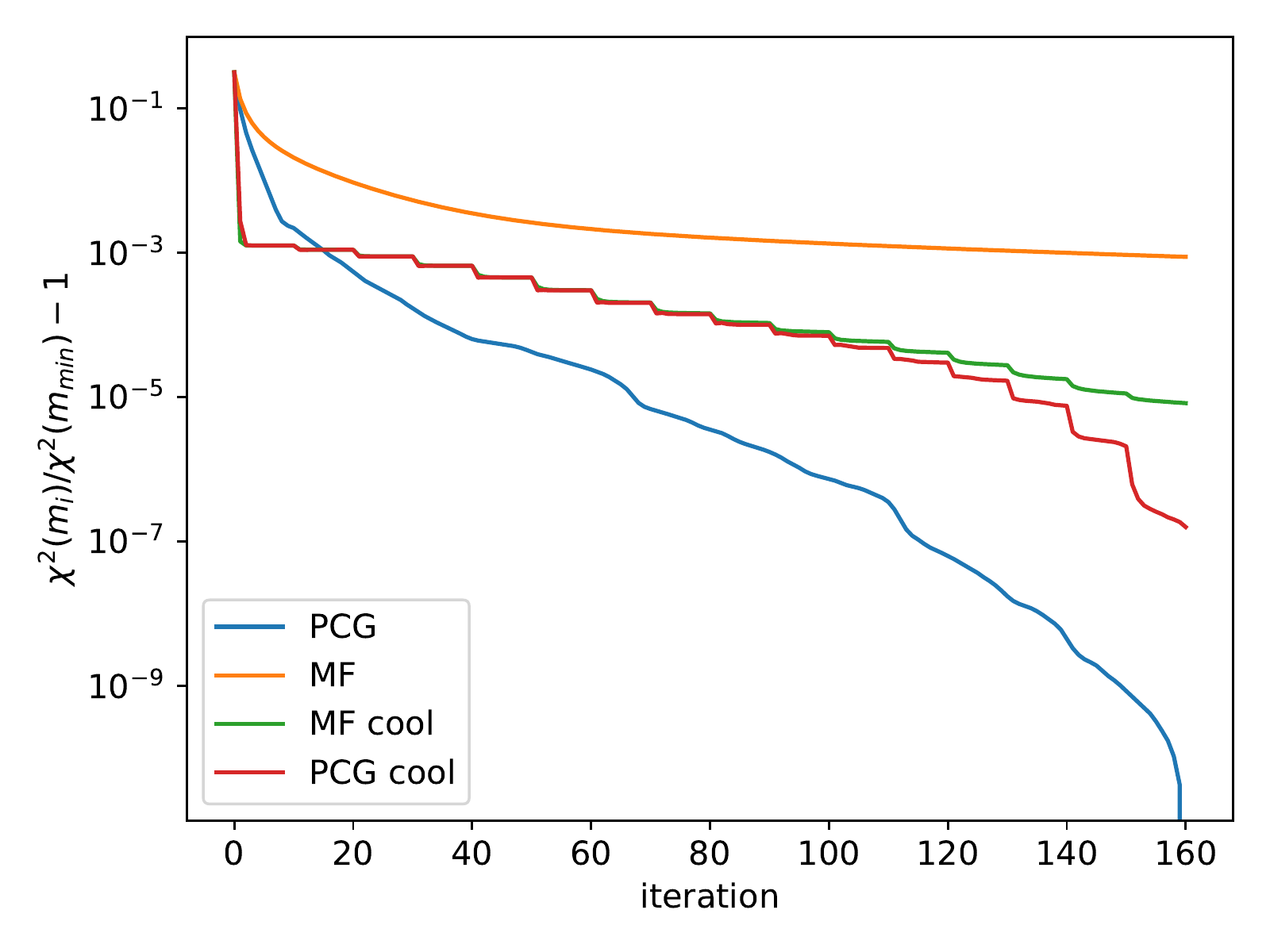}\\
\hskip -0.1truecm
\includegraphics[width=0.335\linewidth]{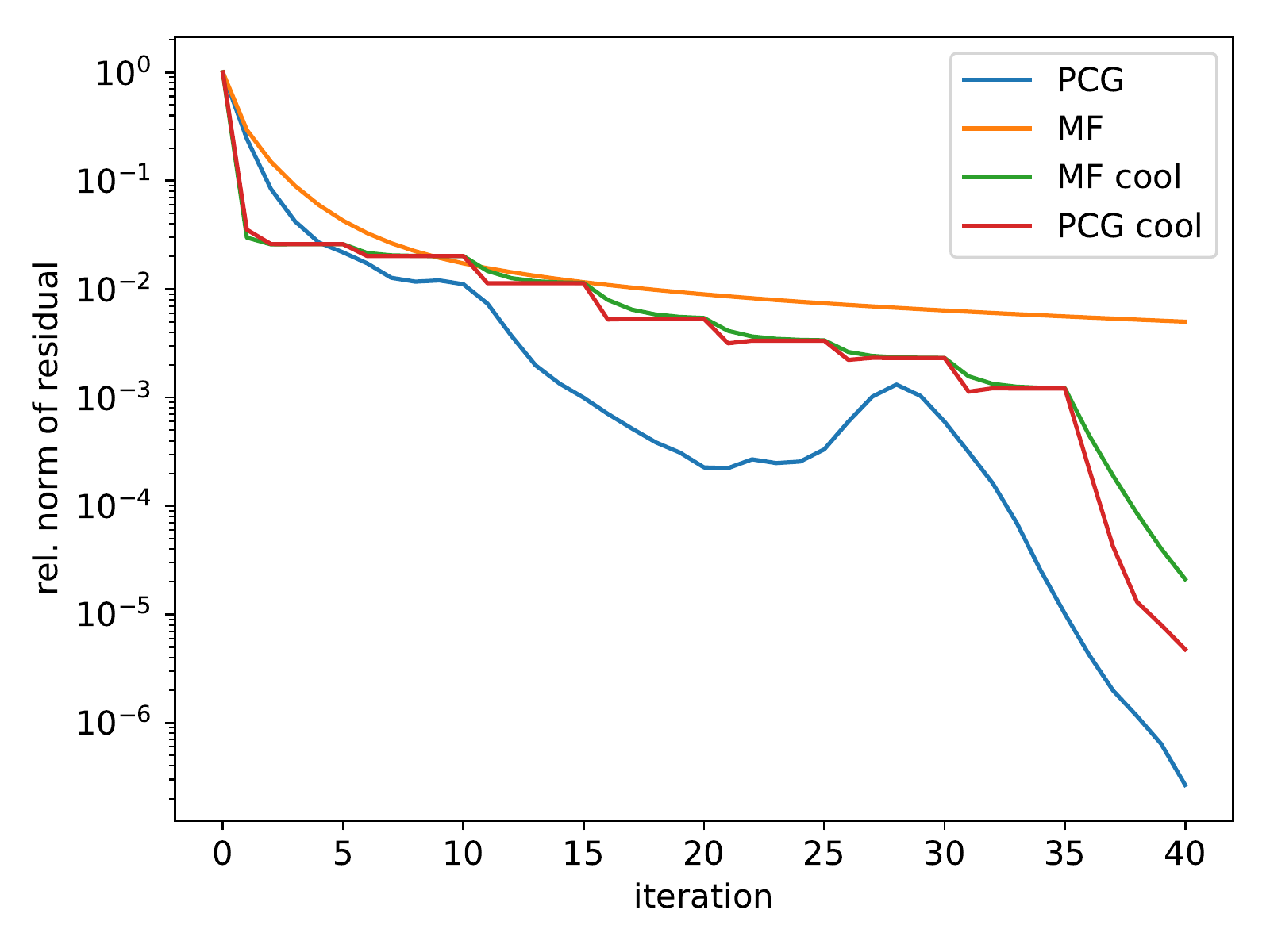}
\hskip -0.1truecm
\includegraphics[width=0.335\linewidth]{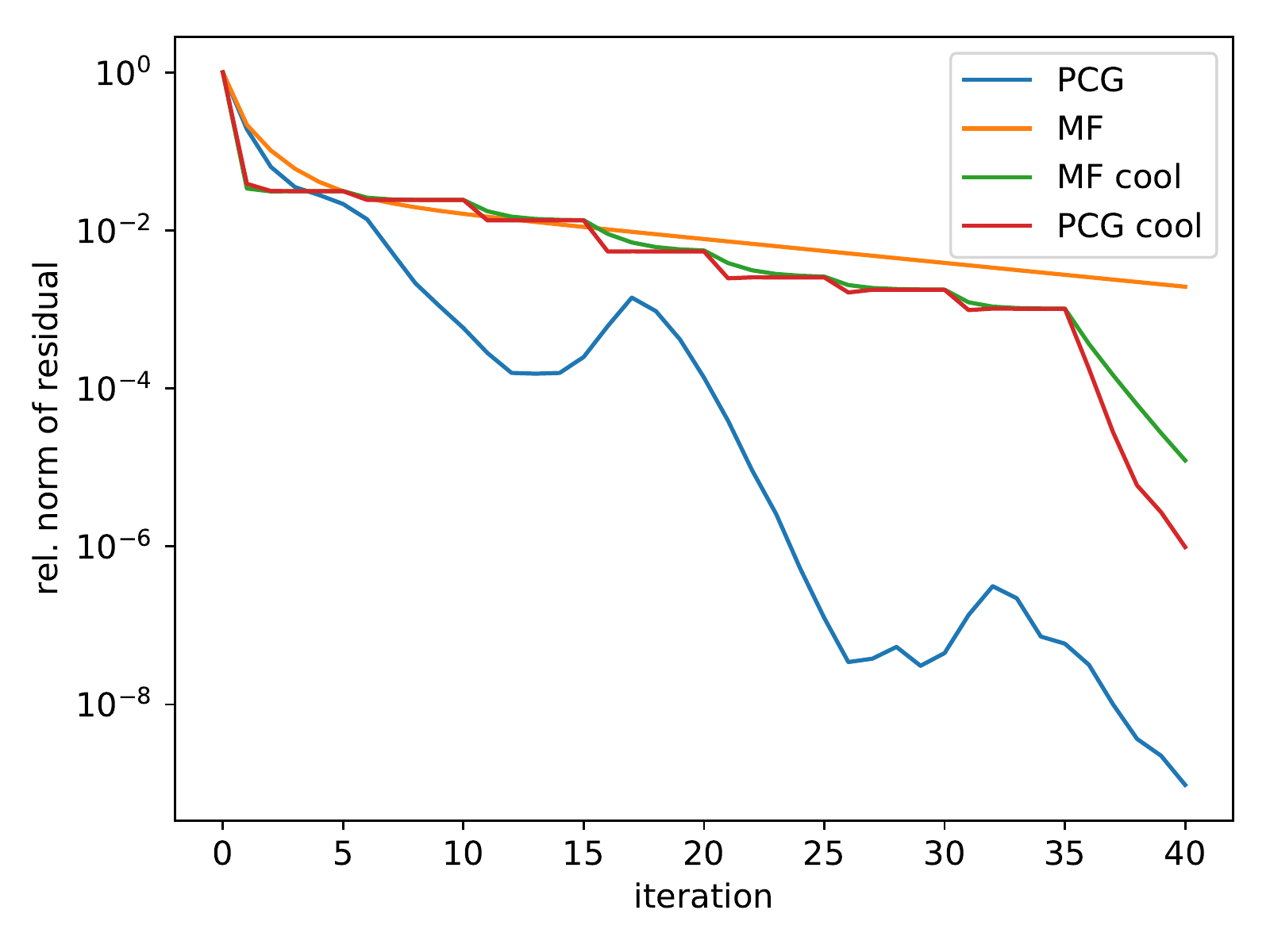}
\hskip -0.1truecm
\includegraphics[width=0.335\linewidth]{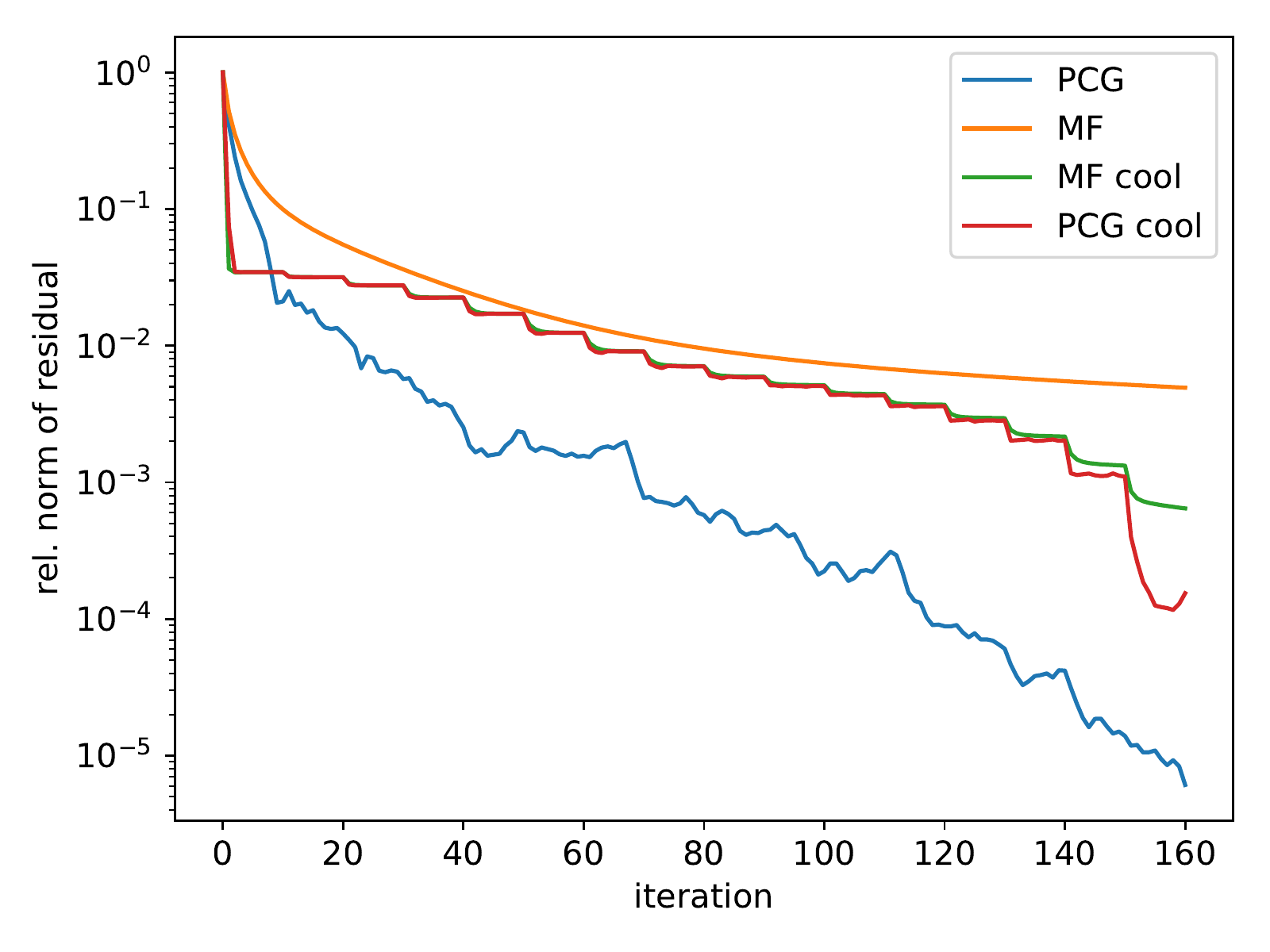}
\caption{Comparison of the convergence for the PCG, the messenger-field methods stand-alone and incorporated within a cooling scheme, for the first, second, and third simulated data sets (left to right,
respectively), assuming a low noise level. The cooling scheme is $8\times 5$ for the first and second data sets and $16\times 10$ for the third.}
\label{fig:MMalpha100case0_PCGvscool}
\end{figure*}

More recently, \cite{HufNae17}~\citep[see also][]{Huff2018} proposed the application of the messenger-field technique to the map-making problem. Below, we discuss the approach of this former work.
The proposal here is again to split the noise covariance into two parts, $\mathbf{N}  \, \equiv \, \mathbf{\bar N} \, + \, \mathbf{T}$, where $\mathbf{T} \, = \, \tau \mathbf{I}$, $\tau = \min(\mbox{diag}(\mathbf{N}))$. Subsequently using Eq.~(\ref{eqn:invSplit}) we can rewrite the system matrix of the map-making equation, Eq.~(\ref{eqn:mapMake}), as
\begin{eqnarray}
\mathbf{P}^{\, t} \, \mathbf{N}^{-1}\, \mathbf{P} \, = \, \mathbf{P}^{\, t} \, \mathbf{T}^{-1}\, \mathbf{P} \, - \, \mathbf{P}^{\, t} \, \mathbf{T}^{-1}\,\big(\mathbf{\bar{N}}^{-1}\,+\,\mathbf{T}^{-1}\big)^{-1} \, \mathbf{T}^{-1} \,  \mathbf{P},
\end{eqnarray}
where the first term on the right-hand side corresponds to matrix $\mathbf{C}$ and the second one to matrix $\mathbf{D}$ as defined in Eq.~(\ref{eqn:matSplit}).
Following the formalism from Sect.~\ref{sect:basicsMF} we can now write the messenger-field equations for this system, which read
\begin{eqnarray}
\big(\mathbf{\bar{N}}^{-1}\,+\,\mathbf{T}^{-1}\big) \, \mathbf{t}_{\phantom{ML}} & = & \mathbf{\bar{N}}^{-1}\,\mathbf{d}\,+\,\mathbf{T}^{-1}\,\mathbf{P}\,\mathbf{m}_{ML},\\
\big(\mathbf{P}^{\, t} \, \mathbf{T}^{-1}\, \mathbf{P}\big) \, \mathbf{m}_{ML} & = & \mathbf{P}^{\, t}\,\mathbf{T}^{-1}\,\mathbf{t},
\end{eqnarray}
with the messenger field~$\mathbf{t}$ appearing explicitly, or
\begin{eqnarray}
\big[\mathbf{I} \, - \, \big(\mathbf{P}^{\, t} \, \mathbf{T}^{-1}\, \mathbf{P}\big)^{-1} \, \mathbf{P}^{\, t} \, \mathbf{T}^{-1}\,\big(\mathbf{\bar{N}}^{-1}\,+\,\mathbf{T}^{-1}\,\big)^{-1}\,\mathbf{T}^{-1}\,\mathbf{P}\big] \, \mathbf{m}_{ML} \, = \, \nonumber\\
 = \, \big(\mathbf{P}^{\, t} \, \mathbf{T}^{-1}\, \mathbf{P}\big)^{-1}\,\mathbf{P}^{\, t}\,\mathbf{N}^{-1}\, \mathbf{d},
\end{eqnarray}
without it. We note that unlike in Eq.~(\ref{eqn:mFieldDefGen}) the matrix $\mathbf{P}^{\, t}$ is taken out of the definition of the messenger field. Solving any of these two sets of equations using fixed-point iterations is equivalent to the messenger-field solver. For comparison we also solve the last equation using the CG technique. The latter is equivalent to solving the map-making equation, Eq.~(\ref{eqn:mapMake}), using a PCG method with the preconditioner taken to be $\mathbf{M} \, =\,  \mathbf{P}^{\, t} \, \mathbf{T}^{-1}\, \mathbf{P}$, which in the case under consideration is equivalent to the standard preconditioner.

We note, following~\citet{HufNae17}, that if the $i$th approximation issued by the fixed point method is unbiased, that is, $\langle \mathbf{x}^{\left(i\right)}\,-\,\mathbf{m}\rangle \, = \, 0$, where $\langle \dots \rangle$ denotes an average over noise realizations, then all the subsequent approximations will also be unbiased. In particular, if the initial guess is chosen to be an unbiased (e.g., simple binned, Eq.~\eqref{eq:MMmodifyinit}) estimate of the sky signal, then all the following up estimates will be unbiased and the entire point of the iterations will be to converge on estimates with minimal statistical uncertainty. This is unlike the case of the PCG, where both statistical and systematic uncertainties are simultaneously improved on during the iterations.

\subsection{Simulated data}
\label{sect:simsMapMake}

We simulate mock time-ordered data $\mathbf{d}$ as a sum of two terms, one corresponding to the sky signal and the other to instrumental noise. These are computed as
\begin{eqnarray}
\mathbf{d}_t \, = \, \mathbf{I}_{p\left(t\right)} \, + \mathbf{Q}_{p\left(t\right)}\,\cos 2\varphi(t)\, + \mathbf{U}_{p\left(t\right)}\,\sin 2\varphi(t) + \mathbf{n}_t,
\end{eqnarray}
where $p(t)$ denotes the sky pixels observed at time $t$ and $\varphi(t)$ is the corresponding orientation of the polarizer.
The signal terms are read off from the signal-only maps of Stokes parameters $I$, $Q$, and $U$, following the assumed scanning strategy defined by $p(t)$ and $\varphi(t)$. These maps are produced in the Healpix pixelization with the resolution parameter $n_{side}$ set to $1024$. These signals are random realizations of the CMB anisotropies corresponding to the currently preferred cosmological model~\citep{Planck2015}.

\begin{figure*}[!ht]
\centering
\hskip -0.1truecm
\includegraphics[width=0.335\linewidth]{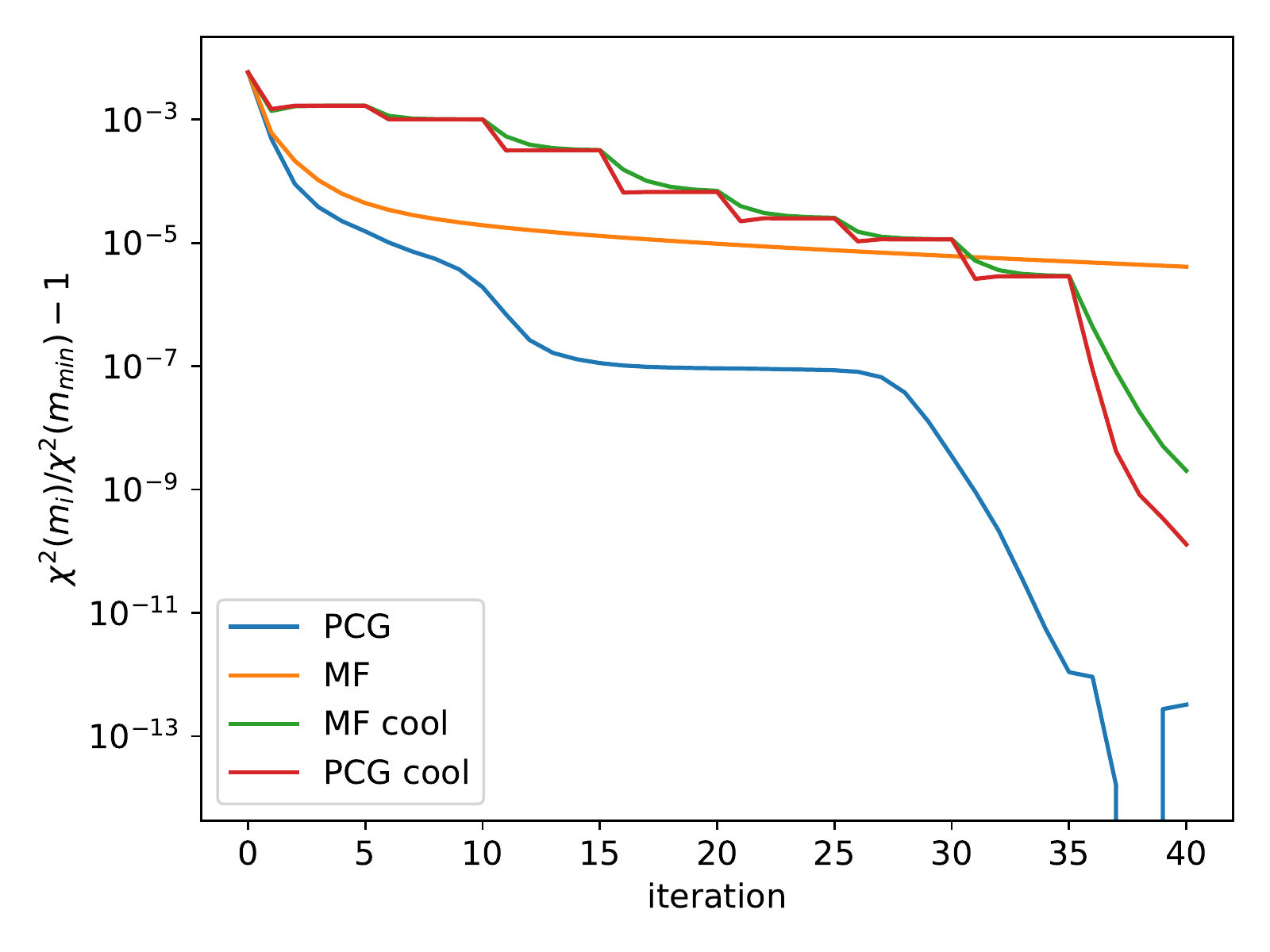}
\hskip -0.1truecm
\includegraphics[width=0.335\linewidth]{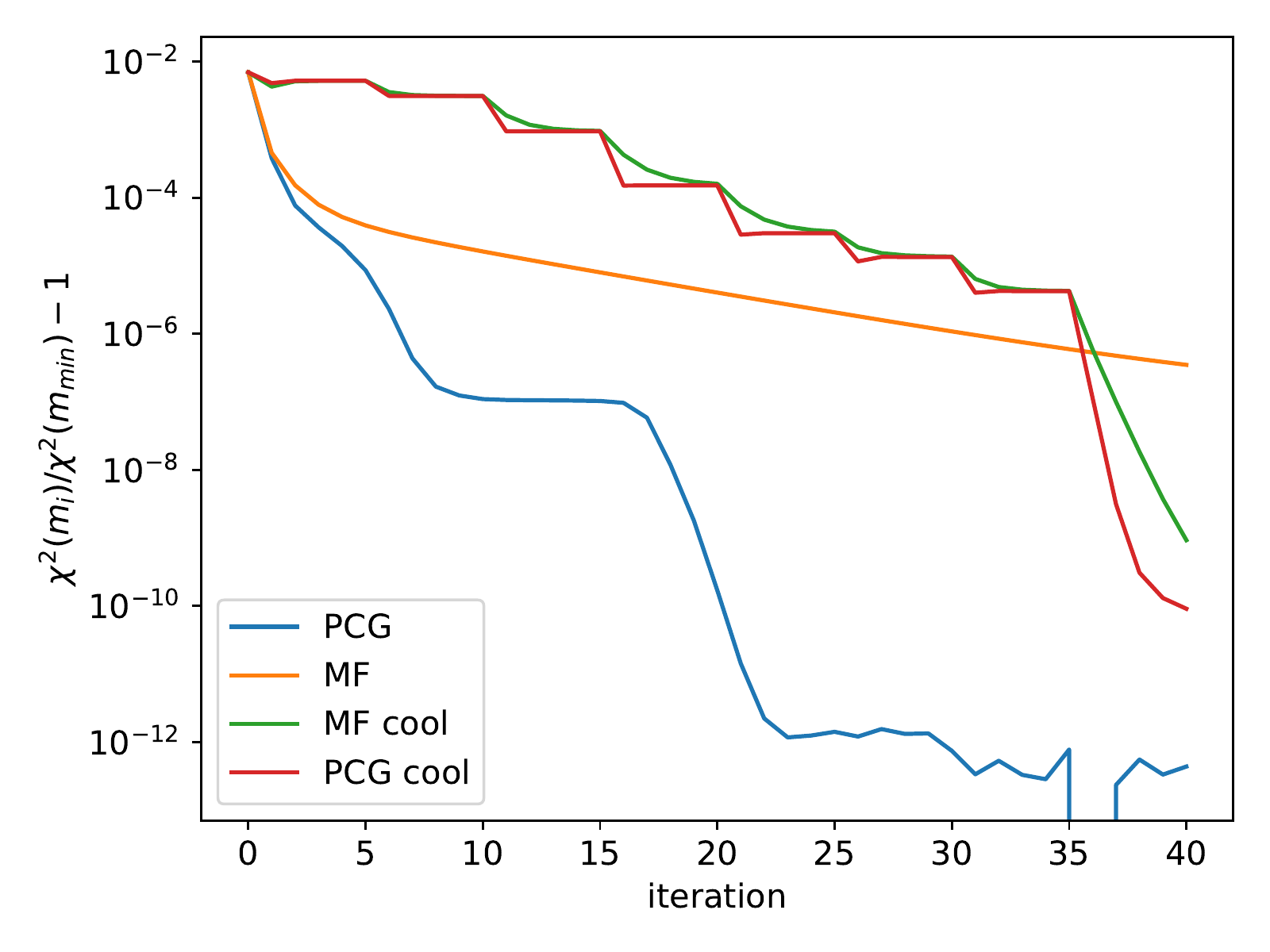}
\hskip -0.1truecm
\includegraphics[width=0.335\linewidth]{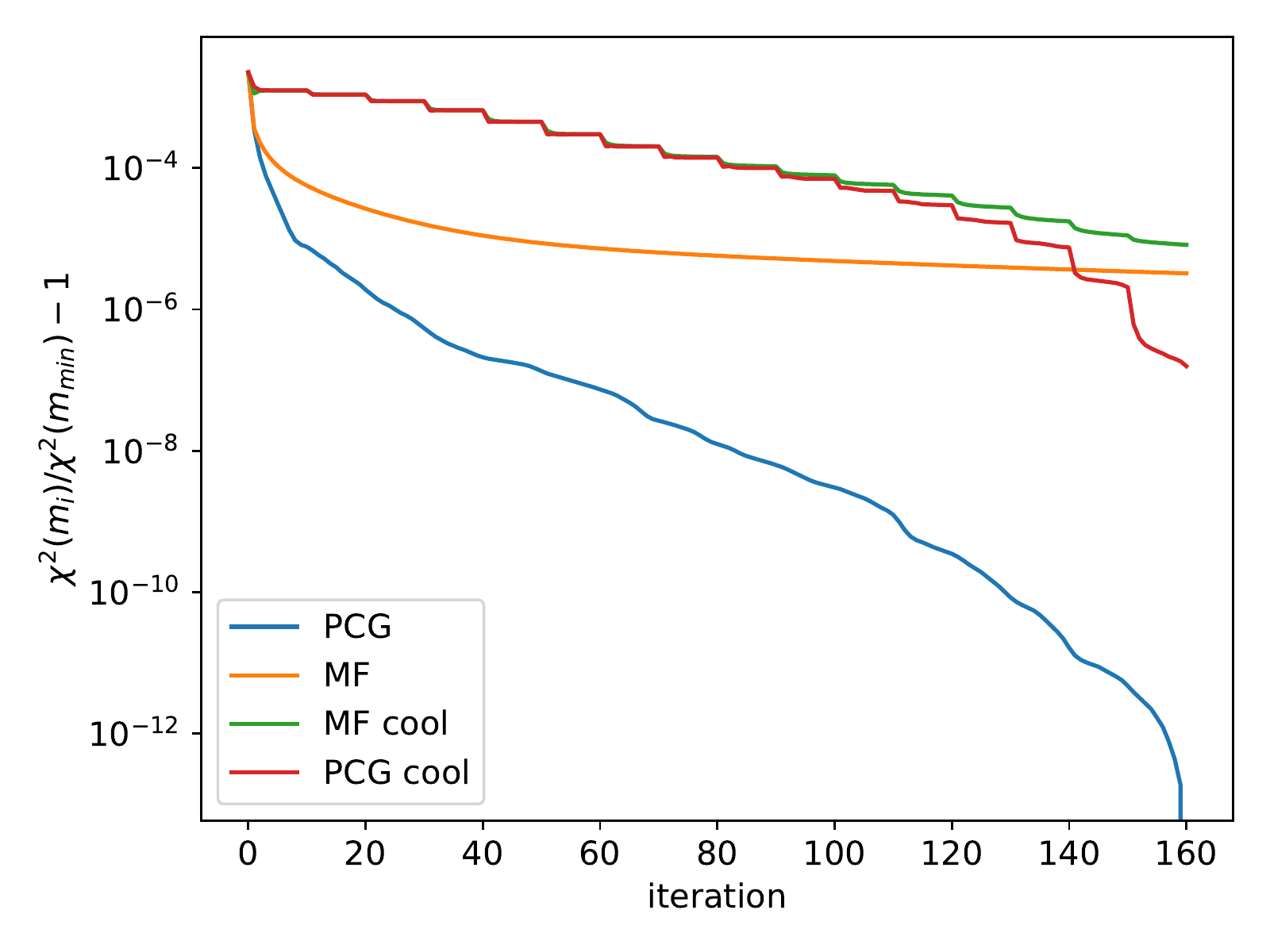}\\
\hskip -0.1truecm
\includegraphics[width=0.335\linewidth]{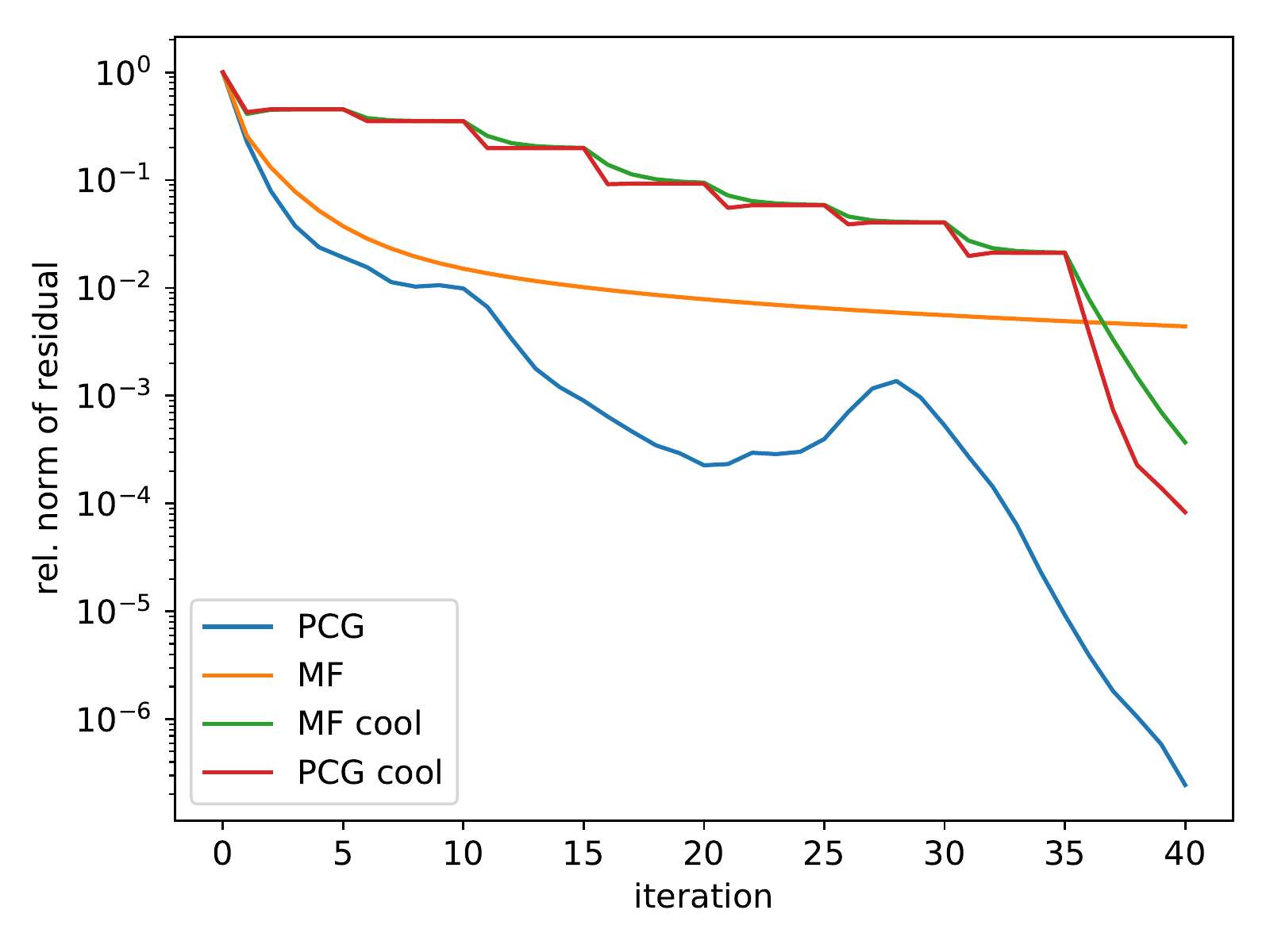}
\hskip -0.1truecm
\includegraphics[width=0.335\linewidth]{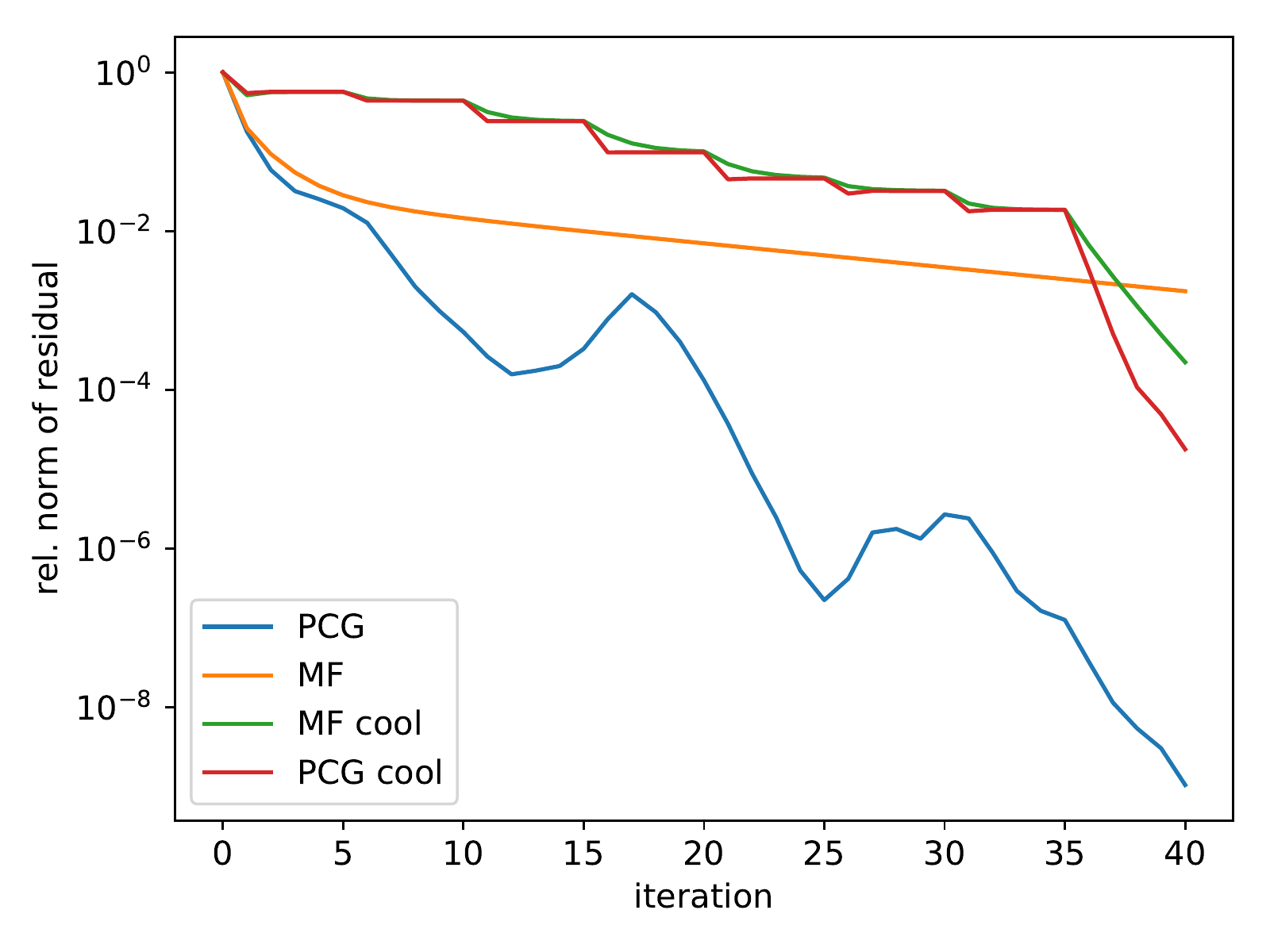}
\hskip -0.1truecm
\includegraphics[width=0.335\linewidth]{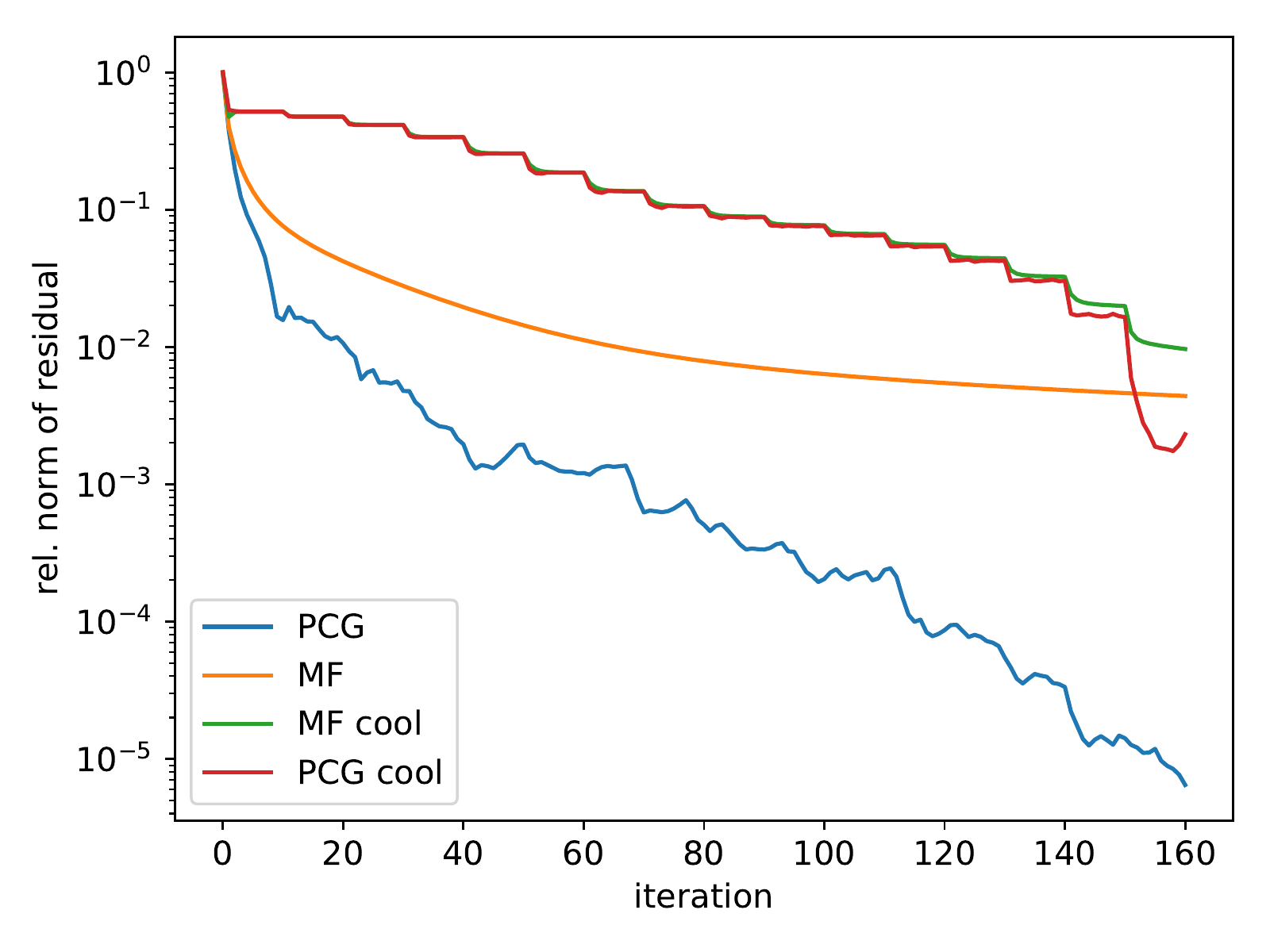}
%
\caption{As in Fig.~\ref{fig:MMalpha100case0_PCGvscool} but for the high noise level.}
\label{fig:MMalpha2000case2_PCGvscool}
\end{figure*}

We produce three data sets with different statistical properties. Each data set comprises the data of a single detector that are however scaled to represent the performance of an entire detector array and thus are more representative of the current data.
In all cases, we assume 
a raster scan pattern in the sky coordinates, when a rectangular sky patch made of $256 \times 256$ Healpix pixels is scanned either horizontally, that is, in right ascension, or vertically,  in declination. The patch is centered at the equator. The length of the simulated data vector is the same and roughly equal to $10^8$.

In the first data set, the sky patch is first scanned horizontally and later vertically.
The horizontal scanning assumes
$256$ complete sweeps (i.e., left-to-right followed by right-to-left), each pixel being sampled on average four times on each sweep. Once this is done, the declination is changed and the new horizontal scan commences. This is repeated $256$ times with each horizontal scan corresponding to a different row of the Healpix pixels.
The vertical scan is implemented in a similar way. 

In this case we assume that the polarizer direction is quickly modulated so the full $2\pi$ angle is sampled within each single crossing of each sky pixel. This is ensured by setting the polarizer angle in the sky coordinates to follow a repeating sequence of $0, \pi/4, \pi/2, 3\pi/4$. In practice, this could mimic the case of an experiment using a smoothly rotating half wave plate.

For the  second data set, we divide it into four equal consecutive subsets, each of which implements the same raster scan made of horizontal scanning within the first half of the subset followed by the vertical scan in the second half. However, the scanning is assumed to be faster and there is only one sample taken per pixel for each pixel crossing. This ensures the same data length. For each subset, the angle of the polarizer in the sky coordinates is fixed and equal to $ 0, \pi/4, \pi/2, 3\pi/4$. This scanning strategy mimics an experiment where the polarizer is stepped discretely only after each of the four subscans.

In the case of the  third data set we progressively change the throw of the scan chop decreasing it gradually to half of the full scan width. We do so for both horizontal and vertical subscans. The scan speed is assumed fixed and tuned in a way that we obtain four observations of each pixel on a single pixel crossing. 
This produces a deeply observed core region where the number of observations per pixels can be as much as three orders of magnitude higher than the number of observations of the outer pixels.
We also assume a smooth polarization angle rotation with the rotation speed fixed in such a way that the polarizer angle changes by $22.5$ degrees on a single pixel crossing. 

This scan strategy is the most realistic from the three considered here reflecting the inhomogeneity of the sky sampling and allowing for imperfect sampling of the polarization angles per pixel. Consequently, for the scan parameters adopted in our simulations, the $3\times 3$ blocks of the block-diagonal preconditioner display a range of condition numbers from $2$ (perfect sampling) to over $20$. The overall condition number of the block-diagonal preconditioner, which accounts for both sky and angle sampling inhomogeneities, is equal to $\sim 1.5\times 10^4$. 

We simulate the instrumental noise as a correlated noise with a power spectrum given by
\begin{eqnarray}
{\mathcal P}(f) \, \equiv \, \sigma^2\,(1+\frac{f_{knee}}{f}),
\end{eqnarray}
where $1/f_{knee}$ is taken to be approximately $500$ times longer than the sampling rate, corresponding to the length of a single full sweep of the sky. We further apodize the low frequency noise effectively flattening the noise spectrum for frequencies lower than a tenth of the knee frequency.

We consider two different noise levels, one ensuring a relatively high signal-to-noise ratio (S/N) of the resulting maps, with rms noise $\sigma_{rms}^{Q,U} \, = \sigma_{rms}^I\cdot\sqrt{2} \simeq 2 \mu$K for the recovered $Q$ and $U$ maps, and the other with lower S/N, corresponding to $\sigma_{rms}^{Q,U} \simeq 30 \mu$K. We refer to these cases as the low- and high-noise data.

We note, that if the instrumental noise were white then the two first scanning strategies would have been equivalent and the standard, block-diagonal preconditioners in both these cases would have been identical.
This is however not the case in the presence of the correlated noise. In fact we expect that the off-diagonal noise correlation of the recovered $Q$ and $U$ maps should be small for the first data set with the quickly rotating HWP while they should be non-negligible in the case of the second data set with the stepped polarizer, leading to different convergence patterns of the studied solvers.

\subsection{Numerical results}
\label{sect:numResMapMake}

\begin{figure*}[!ht]
\centering
\includegraphics[width=0.385\linewidth]{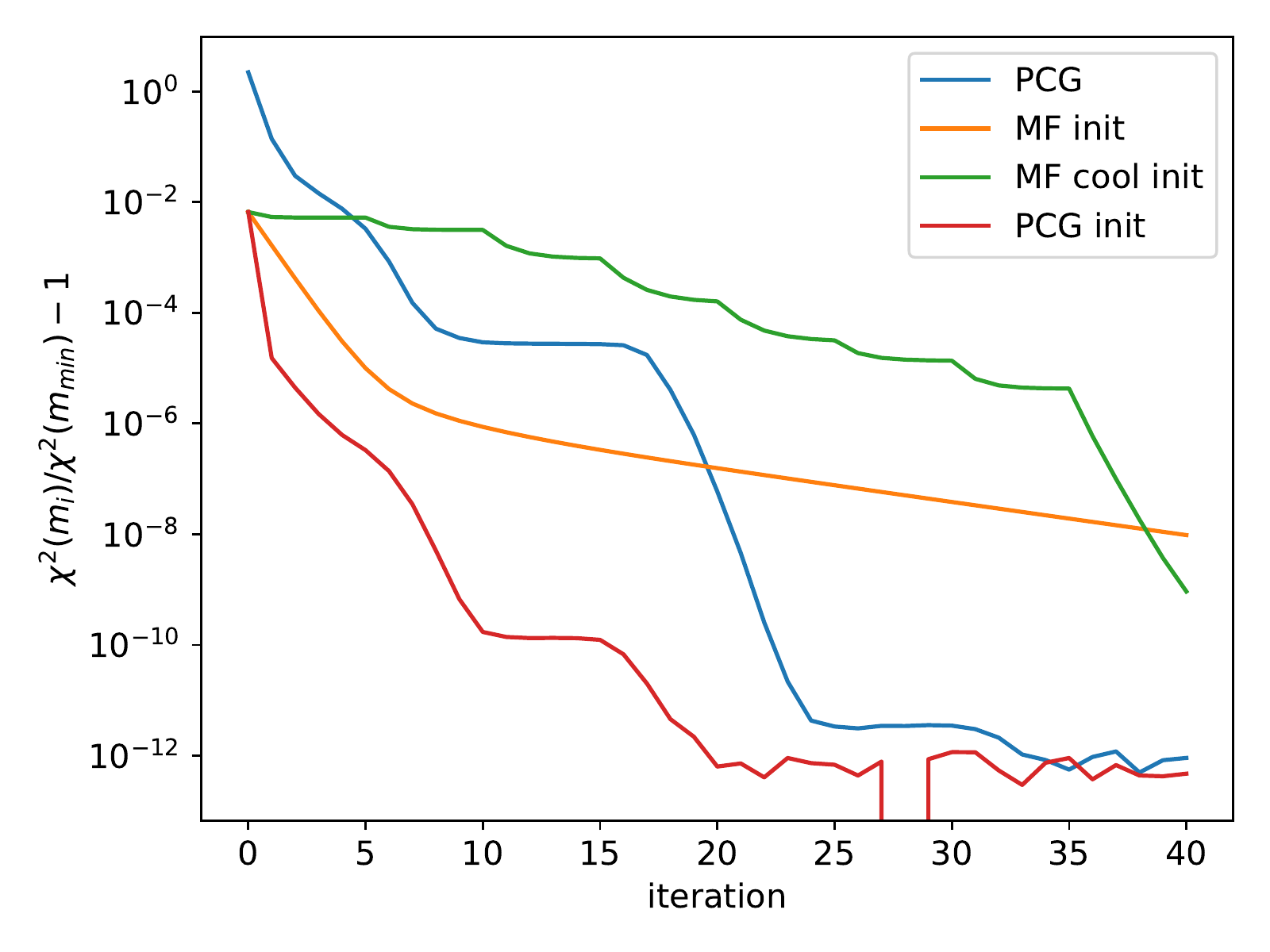}
\includegraphics[width=0.385\linewidth]{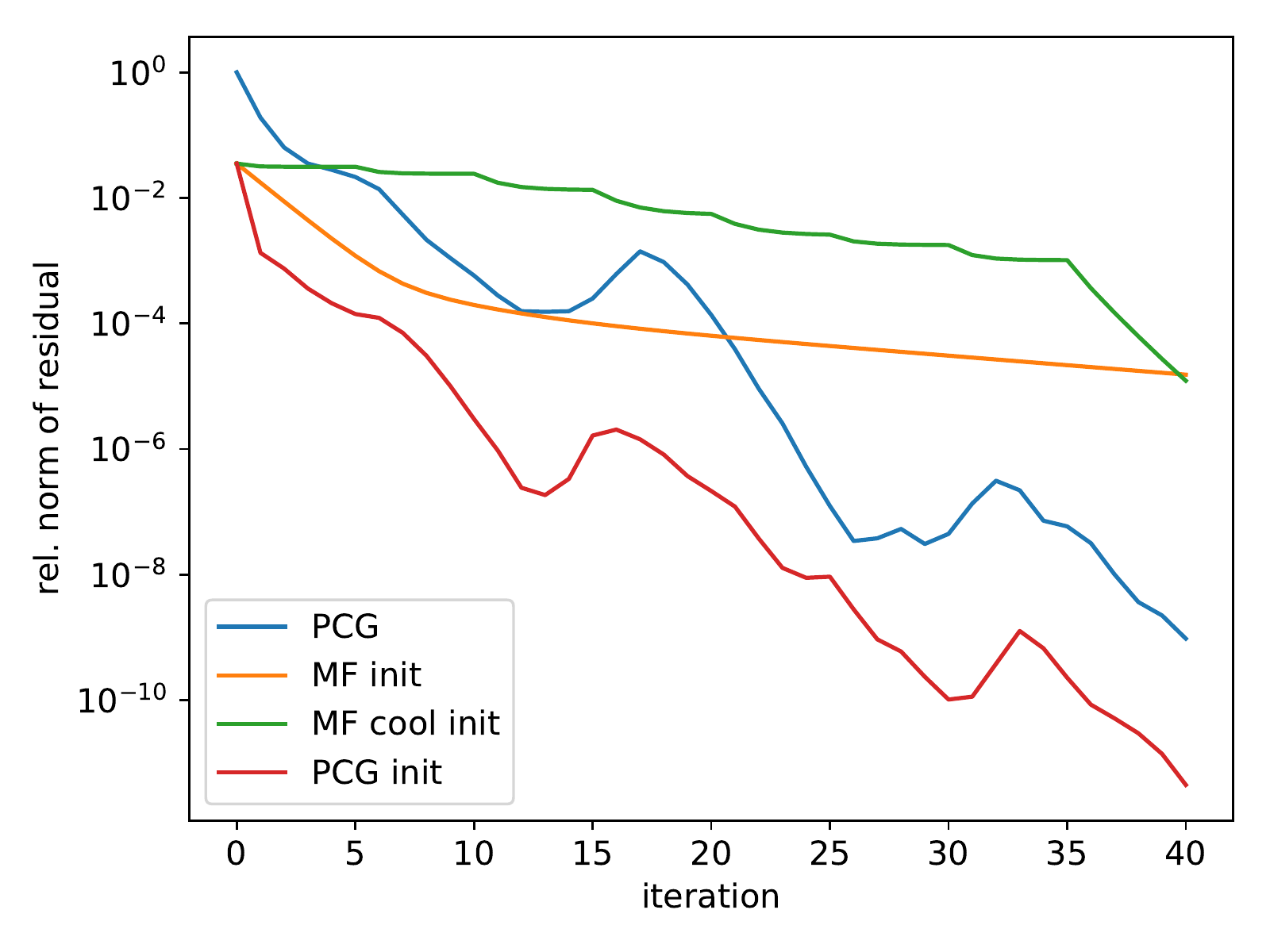}
\caption{Comparison of the convergence rates of different iterative solvers for a non-zero starting vector, $\mathbf{m}^{(0)}$, as given in Eq.~(\ref{eq:MMmodifyinit}). For comparison the blue curve shows the case of the PCG with a starting vector of zeros. The results are for the second scanning strategy and the low-noise case.}
\label{fig:MMalpha100case2_PCGvsPCGinit}
\end{figure*}

\subsubsection{Convergence metric}
\label{sect:mmmetric}

For measuring the error of an approximation~$\mathbf{x}$ we consider, following~\cite{HufNae17}, a $\chi$-measure
\begin{equation}
\label{eq:MMerrormeasure}
        \chi^2(\mathbf{x}) \, = \, (\mathbf{d} - \mathbf{P}\mathbf{x})^{\,t}\,\mathbf{N}^{-1}(\mathbf{d} - \mathbf{P}\mathbf{x}).
\end{equation}
Analogously to the Wiener-filtering application, this measure is minimized by the maximum-likelihood estimate (\ref{eqn:mapMake}) and is equivalent to the energy norm of the error, Eq.~\eqref{eq:energynorm}, with respect to the system matrix, $\mathbf{A} \, \equiv \,  \mathbf{P}^{\, t} \, \mathbf{N}^{-1}\, \mathbf{P}$. Indeed,
\begin{eqnarray}
\chi^2(\mathbf{x}) & = & \| \mathbf{x} \, - \, \mathbf{m}_{ML}\|_{\;\mathbf{P}^{\, t} \, \mathbf{N}^{-1}\, \mathbf{P}} \, + \, \nonumber \\
& + & \mathbf{d}^t\,(\mathbf{N}^{-1} \, - \, \mathbf{N}^{-1} \mathbf{P} \, (\mathbf{P}^t\,\mathbf{N}^{-1}\,\mathbf{P})^{-1}\, \mathbf{P}^t\,\mathbf{N}^{-1})\,\mathbf{d},
\end{eqnarray}
and thus,
\begin{eqnarray}
\chi^2(\mathbf{m}_{ML}) & = & \mathbf{d}^t\,(\mathbf{N}^{-1} \, - \, \mathbf{N}^{-1} \mathbf{P} \, (\mathbf{P}^t\,\mathbf{N}^{-1}\,\mathbf{P})^{-1}\, \mathbf{P}^t\,\mathbf{N}^{-1})\,\mathbf{d}.
\label{eqn:chi2mlMap}
\end{eqnarray}
As before this value is not directly available. However, we can compute the average value of $\chi^2(\mathbf{m}_{ML})$ over the statistical ensemble of the input data realizations and use it as a benchmark for the convergence using the $\chi^2$-measure. This can be done analytically, observing that the matrix on the right-hand side of Eq.~\eqref{eqn:chi2mlMap} is a projection operator, which projects out all time-domain modes, which are sky stationary, that is, they are objects of the form $\mathbf{P}\,\mathbf{y}$ for some arbitrary pixel-domain object $\mathbf{y}$. If so,
\begin{eqnarray}
\chi^2(\mathbf{m}_{ML})& = & \mathbf{n}^t\,(\mathbf{N}^{-1} \, - \, \mathbf{N}^{-1} \mathbf{P} \, (\mathbf{P}^t\,\mathbf{N}^{-1}\,\mathbf{P})^{-1}\, \mathbf{P}^t\,\mathbf{N}^{-1})\,\mathbf{n},
\label{eqn:chi2mlMap_v2}
\end{eqnarray}
and
\begin{eqnarray}
\langle\chi^2(\mathbf{m}_{ML})\rangle_{noise}  & = & {\rm tr}\,(\mathbf{I} \, - \, \mathbf{N}^{-1/2} \mathbf{P} \, (\mathbf{P}^t\,\mathbf{N}^{-1}\,\mathbf{P})^{-1}\, \mathbf{P}^t\,\mathbf{N}^{-1/2})\nonumber\\
& = & n_t \, - \, n_{Stokes}\,n_{pix}\, \equiv \, n_{DOF},
\label{eqn:chi2mlMapAv}
\end{eqnarray}
where $n_t$ and $n_{pix}$ denote the sizes of the data set in time and pixel domains, respectively, and assuming that the system matrix, $\mathbf{P}^{\, t} \, \mathbf{N}^{-1}\, \mathbf{P}$, is non-singular, and considering that
$\mathbf{N}^{-1/2}\,\langle \mathbf{n}\,\mathbf{n}^t\rangle \, \mathbf{N}^{-1/2} \, = \, \mathbf{I}$.
With this value, we then define the convergence criterion in terms of the $\chi^2$-measure by requiring that the incremental change of $\chi^2$ between two consecutive iterations is sufficiently small as compared to $n_{DOF}$.

We note that in the figures, in order to make the behavior of the $\chi$-measure more conspicuous, instead of the $\chi^2$ itself, we plot its relative difference with respect to the minimal value of $\chi^2$ derived within the PCG iterations,
which we denote as $\chi^2(\mathbf{m}_{min})$. The plotted quantity is then given by,
\begin{equation}
\label{eq:MMrelchinorm}
        \chi^2(\mathbf{x})/\chi^2(\mathbf{m}_{min}) - 1.
\end{equation}
For completeness we also plot the standard relative residual defined as
\begin{equation}
\label{eq:MMrelres}
        \frac{\| \mathbf{P}^{\,t}\,  \mathbf{N}^{-1} \mathbf{d} - \big(\mathbf{P}^{\, t} \, \mathbf{N}^{-1}\, \mathbf{P}\big) \, \mathbf{x}\|}
    {\| \mathbf{P}^{\,t}\,  \mathbf{N}^{-1} \mathbf{d}\|}.
\end{equation}

\subsubsection{Performance}

We reconstruct the sky signal from the data using different solvers as discussed here and compare their relative performances. In all cases, we use the same pixelization for the recovered maps as we used for the simulations, that is, Healpix with $n_{side}=1024$. We validated our implementation by running cases with noise-free data and recovering the input maps within the numerical precision. We also found that the results produced by different solvers for each of the data sets agree. 

We first compare the convergence of the PCG solver, of the MF iterations without cooling, and of the PCG and MF with the $8\times 5$ cooling scheme. The results for the low-noise and high-noise cases are given separately in Figs.~\ref{fig:MMalpha100case0_PCGvscool} and 5.

We can see that as claimed in \cite{HufNae17} the MF with cooling technique indeed reaches higher accuracy in comparison to MF without cooling. However, the standard PCG is in these experiments still superior. As in the Wiener-filter application (Sect.~\ref{sect:numResWF}), we observe that in the low-noise cases the cooling technique, used either with PCG or with MF, improves more rapidly on the solution within the first iterations than the PCG method with no cooling. We attribute this to the fact that during those initial iterations the cooling method solves a modified system of the initial equations with an assumed large value of $\lambda$. The approximate solution derived on these first iterations is then equivalent to a simple binned map. This for the low noise cases provides a good rendition of the sky signal, thus leading to an abrupt decrease of the residuals. In the absence of cooling the PCG technique initiated with vectors of zero needs to perform at least a few iterations to reach a comparably good solution. We can however improve on the performance of the stand-alone PCG by using a simple binned map, given by
\begin{equation}
\label{eq:MMmodifyinit}
        \mathbf{m}^{(0)} = (\mathbf{P}^{\,t} \, \mbox{diag}(\mathbf{N})^{-1}\mathbf{P})^{-1} \mathbf{P}^{\,t} \, \mbox{diag}(\mathbf{N})^{-1} \mathbf{d},
\end{equation}
as the starting vector for the PCG solver. Such a map is quickly computable and thus can be readily available at the onset of the solution.

We illustrate these considerations in Fig.~\ref{fig:MMalpha100case2_PCGvsPCGinit}, where we compare the convergence of the PCG run with the initial vector made of zeros and the convergence of the solvers: stand-alone PCG, MF with cooling, and PCG with cooling, assuming $\mathbf{m}^{(0)}$ (E.q.~(\ref{eq:MMmodifyinit})) as the initial vector. The results shown are for the low-noise case.
As expected there is a significant improvement in the overall performance of the PCG method relative to the other solvers but also as compared to the case of the vanishing initial guess. This showcases the importance of the appropriate choice of the initial guess for the PCG approaches in the cases of high-S/N solutions. As the new CMB data sets target predominantly CMB B-mode polarization, the maps of Stokes parameters will increasingly have very high S/N and this observation may be therefore of importance for their data analysis.

We note that all numerical experiments considered here involve non-singular linear systems of equations. If singularities are present then both PCG and MF solvers will typically saturate before reaching the convergence. For the cooled MF this may however not be the case. In particular, if the modified linear systems with $\lambda \simgt 1$ are singularity free then $\lambda$ is effectively a regularization parameter. In such cases the cooled MF may reach the residual level better than the other methods thanks to its ability to adapt amplitudes of the singular modes present in the solution. This however does not change the fact that if these modes are truly singular then their true amplitudes cannot ever be recovered. If the regularization is the appropriate approach to adapt in a given application, then this could also be done in the case of the other solvers.

\section{Conclusions.}
\label{sect:conclusions}

We have shown that the messenger-field solvers of sets of linear equations perform fixed-point iterations of an appropriately preconditioned system of equations. Consequently, in general they are expected to display inferior performance to that of a conjugate gradient solver applied to the same preconditioned systems or, equivalently, to that of a PCG solver with the same preconditioner as implicitly used in the messenger-field method in order to precondition the initial problem. We have backed up this contention with analytic arguments and demonstrated it using numerical experiments involving two applications drawn from modern CMB data analysis practice: Wiener filters and the CMB map-making problem. In addition to the basic implementations of the MF method (Sect.~\ref{sect:basicsMF}), we have considered MF solvers combined with the cooling technique~\citep[Sect.~\ref{sect:coolTech},][]{ElsWan13,HufNae17}, and have shown via numerical results that the cooled MF methods with the cooling schemes as proposed in the literature outperform the standard MF approach. However, the PCG solvers with the preconditioner motivated by the MF methods tend to reach convergence the quickest.

We have compared the performance of the studied methods from the perspective of the number of iterations needed to reach convergence. However, our conclusions are expected to also be directly applicable to considerations involving time-to-solution, as the computational cost per iteration incurred in the different methods is found to be roughly comparable.

We therefore conclude that looking towards the future, advanced preconditioning coupled with the conjugate gradient technique offers the most promise as an expeditious solver, ahead of the messenger-field approach. While at this time, the PCG solvers, with the standard block-diagonal  preconditioner (Eq.~(\ref{eq:MMstandardprec})) in the map-making case, and the preconditioner given by $\mathbf{S}^{-1} + \mathbf{T}^{-1}$ (Eq.~(\ref{eq:WFprec})) in the Wiener filter case, with a potentially appropriately adapted initial guess, should outperform the currently proposed messenger-field approaches. We also note that better preconditioners have already been proposed in particular in the map-making context~\citep[e.g.,][]{Grigori2012, Szydlarski2014}. This notwithstanding, the messenger-field approach may be found of interest in some specific applications.

In the context of the PCG methods, we have found that the convergence may be sped up by an appropriate choice of initial vector. While the gain is largely negligible for the cases with a low-S/N solution, it can become significant if the solution is expected to have high-S/N content. We have found this effect particularly relevant for the map-making procedure, where we have shown that the choice of the simple binned map as the initial vector can result in a significant improvement of the map-making solver convergence.

\begin{acknowledgement}
We thank Dominic Beck and Josquin Errard for their help with simulations and insightful discussions and Kevin Huffenberger and Sigurd N{\ae}ss for their comments on the manuscript. We acknowledge use of HEALpy.
The first two authors' work was supported by the NLAFET project as part of European
    Union's Horizon 2020 research and innovation program under grant 671633.
RS acknowledges support of the French National Research Agency (ANR) contract ANR-17-C23-0002-01 (project B3DCMB).
This research used resources of the National Energy Research Scientific Computing Center (NERSC), a DOE Office of Science User Facility supported by the Office of Science of the U.S. Department of Energy under Contract No. DE-AC02-05CH11231.
\end{acknowledgement}

\appendix

\section{Implementation of PCG for Wiener filter.}
\label{sect:Appendix}

In the context of solving the Wiener-filter problem (Eq.~\eqref{eqn:wienerFilterGen}), each step of the fixed-point method (Eq.~\eqref{eqn:mFieldGenSingleIterated}, resp. Eq.~(\ref{eqn:wienerFilterSplitIterated})) requires one direct and one inverse spherical harmonic transforms, which are assumed to be the most time-consuming elements of the solution process. To keep the same number of transforms in each PCG iteration, we cannot apply a first matrix, $\mathbf{A} = \mathbf{S}^{-1} + \mathbf{N}^{-1}$ , and then precondition the residual by $\mathbf{C}^{-1} \equiv (\mathbf{S}^{-1} + \mathbf{T}^{-1})^{-1}$ as is done in one of the standard PCG implementations listed below in Algorithm~\ref{alg:PCG}. This implementation involves two direct and two inverse transforms: in the evaluation of $\mathbf{A} \mathbf{p}^{(i-1)}$ and in $\mathbf{C}^{-1} \mathbf{r}^{(i)}$.

\begin{algorithm} 
\caption{PCG for  $\mathbf{A} \mathbf{s} = \mathbf{b}$ with the preconditioner $\mathbf{C}$}
\label{alg:PCG} 
Given $\mathbf{s}^{(0)}$, $\mathbf{r}^{(0)} = \mathbf{b} - \mathbf{A} \mathbf{s}^{(0)}$, $\widetilde{\mathbf{r}}^{(0)} = \mathbf{C}^{-1} \mathbf{r}^{(0)}$, $\mathbf{p}^{(0)} = \widetilde{\mathbf{r}}^{(0)}$.\\
For $i = 1, 2, \dots$
        \begin{align*}
                \gamma^{(i-1)} &= \frac{(\mathbf{r}^{(i-1)}, \widetilde{\mathbf{r}}^{(i-1)})}{(\mathbf{p}^{(i-1)}, \mathbf{A} \mathbf{p}^{(i-1)})} ,\\
                \mathbf{s}^{(i)} &=\mathbf{s}^{(i-1)}+\gamma^{(i-1)}\mathbf{p}^{(i-1)}\ ,\\
                \mathbf{r}^{(i)} &=\mathbf{r}^{(i-1)}-\gamma^{(i-1)}\mathbf{A} \mathbf{p}^{(i-1)}\ ,\\
                \widetilde{\mathbf{r}}^{(i)} &= \mathbf{C}^{-1} \mathbf{r}^{(i)}\ ,\\
                \delta^{(i)} &= \frac{(\mathbf{r}^{(i)},\widetilde{\mathbf{r}}^{(i)})}{(\mathbf{r}^{(i-1)},\widetilde{\mathbf{r}}^{(i-1)})} ,\\
                \mathbf{p}^{(i)} &=\widetilde{\mathbf{r}}^{(i)}+\delta^{(i)} \mathbf{p}^{(i-1)}\ .
        \end{align*}
\end{algorithm}

Using the formula for $\mathbf{p}^{(i)}$, $\widetilde{\mathbf{r}}^{(i)}$ and the form of the matrix~$\mathbf{A} = \mathbf{C} - \mathbf{D}$, we can write
\begin{align*}
        \mathbf{A} \mathbf{p}^{(i)} &= \mathbf{A} \big(\widetilde{\mathbf{r}}^{(i)} + \delta^{(i)} \mathbf{p}^{(i-1)} \big)  \\
        &= \mathbf{C} \mathbf{C}^{-1} \mathbf{r}^{(i)} - \mathbf{D}     \widetilde{\mathbf{r}}^{(i)} + \delta^{(i)} \mathbf{A} \mathbf{p}^{(i-1)} \\
        &= \mathbf{r}^{(i)} - \mathbf{D}        \widetilde{\mathbf{r}}^{(i)} + \delta^{(i)} \mathbf{A} \mathbf{p}^{(i-1)}.
\end{align*}
Therefore the vector $\mathbf{A} \mathbf{p}^{(i)}$ can be computed recursively without spherical harmonic transforms and the cost of one PCG iteration is the same (in terms of spherical harmonic transforms) as the cost of one iteration of the fixed-point method, Eq.~\eqref{eqn:mFieldGenSingleIterated}.

Another formula that proved in our numerical experiments to be more stable (yet slightly more costly) is to simultaneously evaluate the vectors $\mathbf{C}^{-1} \mathbf{r}^{(i)}$ and $\mathbf{S}^{-1}\mathbf{C}^{-1} \mathbf{r}^{(i)}$ (recall that $\mathbf{A}\mathbf{C}^{-1} \mathbf{r}^{(i)} = (\mathbf{S}^{-1} + \mathbf{N}^{-1})\mathbf{C}^{-1} \mathbf{r}^{(i)}$). This can be done using direct spherical harmonic transform of one vector, $\mathbf{r}^{(i)}$, and inverse spherical harmonic transform of two vectors\footnote{with the typical computational cost significantly smaller than twice the cost of one single inverse spherical harmonic transform.}. We then simply update $\mathbf{A} \mathbf{p}^{(i)} = \mathbf{A}\mathbf{C}^{-1} \mathbf{r}^{(i)} + \delta^{(i)} \mathbf{A} \mathbf{p}^{(i-1)}$.

Moreover, the properties of (P)CG also allow to evaluate the decrease of the $\chi$-measure without computing it explicitly using Eq.~\eqref{eq:errormeasure} (this computation involves $\mathbf{S}^{-1}$ and therefore also direct and inverse spherical harmonic transforms). The evaluation proposed below is numerically stable; see a thorough analysis in~\cite{StrTic05}. There holds
\[
        \| \mathbf{s} - \mathbf{s}^{(0)} \|^2_{\mathbf{A}} = \sum^{i}_{j=1} \gamma_j \,(\mathbf{r}^{(j)},\widetilde{\mathbf{r}}^{(j)}) + \| \mathbf{s} - \mathbf{s}^{(i)} \|^2_{\mathbf{A}}.
\] 
Using the above discussion on the relationship between the energy norm and the $\chi$-measure, we have
\[
        \chi^2(\mathbf{s}^{(0)}) = \sum^{i}_{j=1} \gamma_j \,(\mathbf{r}^{(j)},\widetilde{\mathbf{r}}^{(j)}) + \chi^2(\mathbf{s}^{(i)}).
\]
After computing $\chi^2(\mathbf{s}^{(0)})$ (that is for zero initial approximation~$\mathbf{s}^{(0)} = 0$ equal to $\mathbf{m}^{\,t}\mathbf{N}^{-1}\mathbf{m}$) we can therefore simply evaluate the $\chi$-measure in every PCG iteration using already computed scalar quantities without any additional spherical harmonic transforms.

\section{Proof of convergence of the messen\-ger-field method.}
\label{sect:MFconverges}

In this appendix we prove that the messenger-field method for Wiener filter is (asymptotically) converging.
Following the discussion in Sect.~\ref{sect:Conv}, we prove the convergence by showing that the eigenvalues of $\mathbf{C}^{-1}\mathbf{D}$ are, in the absolute value, smaller than unity.
First, we note that, since $\mathbf{T} = \tau \mathbf{I}$,
\[\mathbf{C}^{-1}\mathbf{D} = 
        \big[ (\mathbf{S}^{-1} + \mathbf{T}^{-1})^{-1} \mathbf{T}^{-1} \big] \big[ (\mathbf{\bar{N}}^{-1} + \mathbf{T}^{-1} )^{-1} \mathbf{T}^{-1} \big]
\]
is given by multiplication of two symmetric matrices (in brackets).

Algebraic manipulations then yield
\begin{eqnarray*}
        (\mathbf{S}^{-1} + \mathbf{T}^{-1})^{-1} \mathbf{T}^{-1} & = & \mathbf{S} \,(\mathbf{S} + \mathbf{T})^{-1}, \\
    (\mathbf{\bar{N}}^{-1} + \mathbf{T}^{-1} )^{-1} \mathbf{T}^{-1} & = & \mathbf{\bar{N}} \, (\mathbf{\bar{N}} + \mathbf{T})^{-1};
\end{eqnarray*}
see also \cite[Eq.~(5)]{ElsWan13}. Since $\mathbf{S}$, $\mathbf{\bar{N}}$ are symmetric positive semidefinite and $\tau > 0$, the eigenvalues of the matrices above are in the interval $[0, 1)$. For a symmetric matrix~$\mathbf{B}$ there holds $\|\mathbf{B}\|_2 = \rho(\mathbf{B})$.

Finally,
\[
        \rho(\mathbf{C}^{-1}\mathbf{D}) \leq \| \mathbf{C}^{-1}\mathbf{D} \|_2 \leq \| \mathbf{S} (\mathbf{S} + \mathbf{T})^{-1}\|_2 \cdot \|\mathbf{\bar{N}} (\mathbf{\bar{N}} + \mathbf{T})^{-1} \|_2 < 1.
\]
In order to present a close relationship of the derivation of the messenger-field with the Schur complement methods, we present an alternative proof below. An analogous derivation can be used also for proving the convergence of MF in the map-making application.

We start by rewriting Eq.~\eqref{eqn:wienerFilterMessFieldIter} as the system 
\begin{eqnarray*}
\underbrace{\begin{pmatrix}
\mathbf{\bar{N}}^{-1} + \mathbf{T}^{-1} & -\mathbf{T}^{-1} \\
-\mathbf{T}^{-1} & \mathbf{S}^{-1} + \mathbf{T}^{-1}
\end{pmatrix}}_{\displaystyle \equiv \mathbb{A.}}
\begin{pmatrix}
\mathbf{t} \\
\mathbf{s}_{WF}
\end{pmatrix}
=
\begin{pmatrix}
\mathbf{\bar{N}}^{-1}\,\mathbf{m} \\
0
\end{pmatrix}
\end{eqnarray*}
The reduced system (after the elimination of the messenger field $\mathbf{t}$) then corresponds to forming the Schur complement $\mathbb{S}$ of $\mathbb{A}$,
\[
        \mathbb{S} \equiv \big(\mathbf{S}^{-1} + \mathbf{T}^{-1}\big) - \mathbf{T}^{-1} \big(\mathbf{\bar{N}}^{-1} + \mathbf{T}^{-1} \big)^{-1} \mathbf{T}^{-1},
\]
and solving
\[
        \mathbb{S} \mathbf{s}_{WF} = \mathbf{T}^{-1} \big(\mathbf{\bar{N}}^{-1} + \mathbf{T}^{-1} \big)^{-1} \mathbf{\bar{N}}^{-1}\,\mathbf{m}.
\]
The MF iterations are obtained by multiplying (preconditioning) the above system by $(\mathbf{S}^{-1} + \mathbf{T}^{-1})^{-1}$ from the left.

Now we show the bounds on the eigenvalues of 
\[ 
        \big(\mathbf{S}^{-1} + \mathbf{T}^{-1}\big)^{-1} \mathbf{T}^{-1} \big(\mathbf{\bar{N}}^{-1} + \mathbf{T}^{-1} \big)^{-1} \mathbf{T}^{-1} = \mathbf{I} - \big(\mathbf{S}^{-1} + \mathbf{T}^{-1}\big)^{-1} \mathbb{S}.
\]
Since $\mathbb{A}$ is an SPD matrix, its Schur complement $\mathbb{S}$ is also SPD. Moreover, the spectrum of $(\mathbf{S}^{-1} + \mathbf{T}^{-1})^{-1}\mathbb{S}$ satisfies
\[
 \Lambda\left(\big(\mathbf{S}^{-1} + \mathbf{T}^{-1}\big)^{-1}\mathbb{S}\right) 
    = \Lambda\left(\big(\mathbf{S}^{-1} + \mathbf{T}^{-1}\big)^{-1/2}\,\mathbb{S}\,\big(\mathbf{S}^{-1} + \mathbf{T}^{-1}\big)^{-1/2}\right),
\]
and therefore the eigenvalues of $(\mathbf{S}^{-1} + \mathbf{T}^{-1})^{-1}\mathbb{S}$ are positive. Plugging into the above equation the formula for the Schur complement $\mathbb{S}$, we have
\begin{multline*}
        \Lambda\left(\big(\mathbf{S}^{-1} + \mathbf{T}^{-1}\big)^{-1}\mathbb{S}\right) \\
    = \Lambda\left( \mathbf{I} - \big(\mathbf{S}^{-1} + \mathbf{T}^{-1}\big)^{-1/2}\big(\mathbf{\bar{N}}^{-1} + \mathbf{T}^{-1} \big)^{-1}\big(\mathbf{S}^{-1} + \mathbf{T}^{-1}\big)^{-1/2}\right) \\
    = 1 - \Lambda\left( \big(\mathbf{S}^{-1} + \mathbf{T}^{-1}\big)^{-1/2}\big(\mathbf{\bar{N}}^{-1} + \mathbf{T}^{-1} \big)^{-1}\big(\mathbf{S}^{-1} + \mathbf{T}^{-1}\big)^{-1/2}\right).
\end{multline*}
The matrix $(\mathbf{S}^{-1} + \mathbf{T}^{-1})^{-1/2}(\mathbf{\bar{N}}^{-1} + \mathbf{T}^{-1} )^{-1}(\mathbf{S}^{-1} + \mathbf{T}^{-1})^{-1/2}$ is symmetric positive semidefinite. Altogether,
\[
        \Lambda\left(\big(\mathbf{S}^{-1} + \mathbf{T}^{-1}\big)^{-1}\mathbb{S}\right) \in (0,1].
\]
Consequently,
\[
        \Lambda\left(\mathbf{C}^{-1}\mathbf{D}\right) = 1 - \Lambda\left(\big(\mathbf{S}^{-1} + \mathbf{T}^{-1}\big)^{-1}\mathbb{S}\right) \in [0,1).
\]

Finally, we show that the asymptotic convergence of the error in the Euclidean norm $\| \epsilon^{\left(i\right)} \|$, which is assured by the fact that $\rho(\mathbf{C}^{-1}\mathbf{D}) < 1$ (see, e.g., \citet[Section 4.2]{SaaBook03}), proves the asymptotic convergence of the error~$ \epsilon^{\left(i\right)}$ in any norm \mbox{$\|\cdot\|_*$}. Here we use the equivalence of norms on finite-dimensional spaces; see, e.g., \citet[Corollary 5.4.6 and Definition 5.4.7]{HorJohBook13}. In particular, given any norm~\mbox{$\|\cdot\|_*$}, there exist positive constants $c_*\,$, $C_*$\,, such that
\[
        c_* \|\mathbf{v}\|_* \leq \|\mathbf{v}\| \leq C_* \|\mathbf{v}\|_*\,, \quad \mbox{for all} \quad \mathbf{v}.
\]
Consequently,
\[
        \| \epsilon^{\left(i\right)} \|_* \leq (c_*)^{-1}\, \|\epsilon^{\left(i\right)}\| \rightarrow 0\,, \quad \mbox{for} \quad i\rightarrow \infty.
\]


\end{document}